\documentclass[11pt,a4paper]{article}
\pdfoutput=1
\usepackage{jheppub}
\usepackage{graphicx}
\usepackage[latin2]{inputenc}
\usepackage{t1enc}
\usepackage{amsmath}
\usepackage{subfigure}
\usepackage{dsfont}
\usepackage{slashed}
\usepackage{multirow}
\usepackage{lscape}
\usepackage{wrapfig}
\usepackage{array}

\usepackage{setspace}

\usepackage{colordvi}
\usepackage{lscape}
\usepackage{rotfloat}
\usepackage{rotating}
\usepackage{amssymb}

\usepackage{cmap}

\newcommand{\dd}{\textmd{d}}
\newcommand{\be}{\begin{equation}}
\newcommand{\ee}{\end{equation}}

\newcommand{\Z}{\mathcal{Z}}
\newcommand{\F}{\mathcal{F}}

\newcommand{\M}{\mathcal{M}}
\newcommand{\D}{\mathcal{D}}

\newcommand{\tr}{\textmd{tr}\,}
\newcommand{\fracp}[2]{\frac{\partial #1}{\partial #2}}
\newcommand{\eBM}{eB \cdot\M}
\renewcommand{\P}{\mathcal{P}}
\newcommand{\QEDb}{b_1}

\newcommand{\Budapest}{E\"otv\"os University, Theoretical Physics, P\'azm\'any P.\ s.\ 1/A, H-1117, Budapest, Hungary.}
\newcommand{\Regensburg}{Institute for Theoretical Physics, Universit\"at Regensburg, D-93040 Regensburg, Germany.}
\newcommand{\Tata}{Tata Institute of Fundamental Research, Homi Bhabha Road, Mumbai 400005, India.}
\newcommand{\Lendulet}{MTA-ELTE Lend\"ulet Lattice Gauge Theory Research Group. P\'azm\'any P.\ s.\ 1/A, H-1117, Budapest, Hungary.}

\title{The QCD equation of state in background magnetic fields}

\renewcommand{\thefootnote}{\fnsymbol{footnote}} 	

\author[a,b]{G.~S.~Bali,}
\author[a]{F.~Bruckmann,}
\author[a,*]{G.~Endr\H{o}di,\note{Corresponding author}}
\author[c,d]{S.~D.~Katz,}
\author[a]{A.~Sch\"afer,}

\affiliation[a]{\Regensburg}
\affiliation[b]{\Tata}
\affiliation[c]{\Budapest}
\affiliation[d]{\Lendulet}

\emailAdd{gunnar.bali@physik.uni-r.de}
\emailAdd{falk.bruckmann@physik.uni-r.de}
\emailAdd{gergely.endrodi@physik.uni-r.de}
\emailAdd{katz@bodri.elte.hu}
\emailAdd{andreas.schaefer@physik.uni-r.de}

\abstract{
We determine the equation of state of 2+1-flavor QCD with physical quark masses, 
in the presence of a constant (electro)magnetic background field 
on the lattice. 
To determine the free energy at nonzero magnetic fields we
develop a new method, 
which is based on an integral over the quark masses up to asymptotically large 
values where the effect of the magnetic field can be neglected. 
The method is compared to other approaches in the literature and found to be 
advantageous for the determination of the equation of state up to large magnetic fields. 
Thermodynamic observables including the longitudinal and transverse pressure, magnetization, 
energy density, entropy density and interaction measure 
are presented for a wide range of temperatures and magnetic fields, and provided 
in ancillary files.
The behavior of these observables confirms 
our previous result that the transition temperature is reduced by the magnetic field. 
We calculate the magnetic susceptibility and permeability, 
verifying that the thermal QCD medium is paramagnetic around and above 
the transition temperature, while we also find evidence for weak diamagnetism 
at low temperatures. 
}


\begin{document}

\maketitle

\renewcommand{\thefootnote}{\arabic{footnote}}

\section{Introduction}

Quantum Chromodynamics (QCD) is the theory of the strong interactions. 
Its most important properties are the confinement of quarks and gluons at low 
energies and asymptotic freedom at high scales. Lattice simulations of QCD 
have unambiguously shown that at zero quark densities these two, fundamentally 
different regimes 
are connected by a smooth crossover-type 
transition~\cite{Aoki:2006we,Bhattacharya:2014ara}. 
This transition -- through which the dominant degrees of freedom change 
from composite objects (hadrons) to colored quarks and gluons -- has several 
characteristics of both theoretical and phenomenological 
relevance. Besides the nature and the (pseudo)critical temperature 
of the transition, one such characteristic is the equation of state (EoS),
which is the fundamental relation encoding the thermodynamic
properties of the system.

In particular, the EoS gives the equilibrium description of QCD matter, in terms of 
relations between thermodynamic observables like the pressure, the energy
density or the entropy density. These observables enter hydrodynamic models 
that are used to describe the time evolution of the quark-gluon plasma (QGP)
produced in heavy-ion collision experiments~\cite{Teaney:2001av,
Kolb:2003dz}. Besides its role in heavy-ion physics, 
the EoS affects the mass-radius relation of neutron 
stars~\cite{Lattimer:2000nx} and enters cosmological models of the early universe,
with implications,
for example, for dark matter candidates~\cite{Hindmarsh:2005ix}. 

Of particluar relevance is the response of the EoS to changes of 
the control 
parameters of the system. These parameters include 
the temperature, the chemical potentials conjugate to conserved 
charges and, in the present case, a background (electro)magnetic field $B=|\mathbf{B}|$. 
External magnetic fields play an important role in the evolution 
of the early universe~\cite{Grasso:2000wj}, in strongly magnetized neutron 
stars~\cite{Duncan:1992hi}
and in non-central heavy-ion collisions, see, e.g., the recent review~\cite{Kharzeev:2013jha}. 
Magnetic fields induce a variety of exciting effects in the thermodynamics of QCD -- 
for example they significantly affect the phase diagram. The first results in this field 
were obtained using low-energy models and effective theories of QCD, see the summary in, e.g.,
Ref.~\cite{Fraga:2012rr}.
The QCD transition has also been studied extensively on the lattice; we refer the 
reader to reviews on the subject in, e.g.,
Refs.~\cite{D'Elia:2012tr,Bali:2013cf,Szabo:2014iqa}.
A very relevant question in this respect has been the dependence of the
transition temperature $T_c$ and of the nature of the transition
on the magnetic field. In this paper we also address this issue.

Our main objective is to determine the QCD EoS around the crossover transition 
at vanishing chemical potentials, for nonzero background magnetic fields. 
To this end we develop a generalization of the so-called integral method~\cite{Engels:1990vr}, 
which relies on an integration in the quark masses up to asymptotically large values
(a similar integration in the quark masses at $B=0$ was also considered in Ref.~\cite{Borsanyi:2010cj}). 
We calculate thermodynamic observables 
including the pressure, the energy density, the entropy density, the interaction measure, 
the magnetization and the susceptibility 
for magnetic fields of up to $eB=0.7\textmd{ GeV}^2$
for a wide range of temperatures $110\textmd{ MeV}<T<300\textmd{ MeV}$, allowing for 
a comparison with the Hadron Resonance Gas (HRG) model and with 
perturbation theory, at low and high temperatures, respectively. 
At high $T$ we demonstrate that several aspects of perturbative QED 
physics are encoded in the EoS observables. 
Furthermore, our results confirm the observation made in Refs.~\cite{Bali:2011qj,Bruckmann:2013oba} 
that the transition region is shifted to lower 
temperatures as $B$ grows. 
Another phenomenological consequence of the magnetic field is that the pressure 
-- 
if defined as the response against a compression at fixed magnetic flux, see precise definition
in Sec.~\ref{sec:thermo} below -- 
becomes anisotropic, and a significant splitting between the components 
parallel and perpendicular to $\mathbf{B}$ is developed.

The change in the EoS due to the magnetic field has further theoretical 
implications. QCD matter may be thought of as a medium 
with either para- or diamagnetic properties. We establish that around and above 
the transition region the magnetization 
is positive and thus the thermal QCD medium behaves as a paramagnet, with a 
magnetic permeability larger than unity. A possible implication 
of this paramagnetism for heavy-ion collisions has been pointed out recently in 
Ref.~\cite{Bali:2013owa}. In addition, we present evidence for the emergence of 
a weakly diamagnetic region at low temperatures due to pions.

This paper is organized as follows. First we discuss thermodynamic relations 
in the presence of the magnetic field from a general point of view in 
Sec.~\ref{sec:thermo}. We proceed by describing the lattice methods that 
are used to determine the EoS in Sec.~\ref{sec:latobs}, with special 
emphasis on the implications of flux quantization and electric charge renormalization. 
Sec.~\ref{sec:results} contains our main results, followed by the 
conclusions in Sec.~\ref{sec:summary}. 

\section{Thermodynamics in an external magnetic field}
\label{sec:thermo}

The fundamental quantity of thermodynamics is the free energy or thermodynamic potential. 
In terms of the partition function $\Z$ of the system it reads $\F=-T \log\Z$. In the presence of an external magnetic field the density $f=\F/V$ of the free energy in a finite spatial volume $V$ 
can be written as~\cite{landau1995electrodynamics}
\be
f = \epsilon - Ts = \epsilon^{\rm total} -T s - eB \cdot\M ,
\label{eq:fundeq}
\ee
where $\epsilon$ is the energy density of the medium, 
$s$ the entropy density and $\M$ the magnetization. 
Without loss of generality, the magnetic field $\mathbf{B}=B\,\mathbf{e}_z$ is taken to 
point in the $z$ direction, and 
for later convenience, $B$ is given in units of the elementary charge $e>0$. 
Note that the total energy of the system, $\epsilon^{\rm total}=\epsilon+\epsilon^{\rm field}$, 
includes the energy of the medium $\epsilon$ as well as the work necessary to maintain the constant external field, 
$\epsilon^{\rm field}=\eBM$~\cite{kittel2004elementary}. The two expressions in Eq.~(\ref{eq:fundeq}) thus correspond to two different conventions for the definition of the energy density. 
The entropy density and the magnetization can be obtained as 
\be
\frac{1}{V}\fracp{\F}{T} = -s,\quad\quad\quad
\frac{1}{V}\fracp{\F}{(eB)} = -\M.
\label{eq:diffrels}
\ee

The corresponding differential relation for the pressure is somewhat more involved. 
Since the magnetic field marks a preferred direction, the pressures $p_i$ in the transverse (perpendicular to $\mathbf{B}$) and in the longitudinal (parallel to $\mathbf{B}$) 
directions may be different. 
In Ref.~\cite{Bali:2013esa} we have shown that this possible anisotropy depends on the precise definition of $p_i$. Writing the volume as the product of linear extents $V=L_x L_y L_z$, 
the pressure components are related to the 
response of the system to compressions along the corresponding directions, i.e.
\be
p_i = -\frac{1}{V} L_i \frac{\partial \F}{\partial L_i}.
\label{eq:pidef}
\ee
In order to unambiguously define $p_i$, we have to specify the trajectory in parameter 
space, along which the partial derivative is evaluated. In Ref.~\cite{Bali:2013esa}, 
we have distinguished between a setup where the magnetic field $B$ is kept fixed 
during the compression (the ``$B$-scheme''), and a setup where the magnetic flux $\Phi=eB \cdot L_xL_y$
is kept fixed (the ``$\Phi$-scheme''). The $B$-scheme results in {\it isotropic} pressures, whereas the 
$\Phi$-scheme gives {\it anisotropic} pressures:
\be
p_x^{(B)}=p_y^{(B)}=p_z,\quad\quad\quad p_x^{(\Phi)}=p_y^{(\Phi)}=p_z-\eBM.
\label{eq:BPHI}
\ee
The difference in the transverse components for the $\Phi$-scheme 
is due to the fact that the compressing force 
in this case also acts against the magnetic field. 
Note that the definition of the pressures as spatial diagonal components 
of the energy-momentum tensor exhibits the $\Phi$-scheme anisotropy~\cite{Ferrer:2010wz}.
This is due to the fact that the energy-momentum tensor is usually defined 
through the variation of the action with respect to the metric at fixed $\Phi$, 
see Ref.~\cite{Bali:2013esa}. 
In contrast to $p_{x,y}$, the longitudinal pressure is independent of the scheme, and in the thermodynamic limit $V\to\infty$ simplifies to
\be
p_z=-f.
\label{eq:pzdef}
\ee

Note that the appropriate scheme to be used depends on the physical situation that one would like to describe. 
In particular, it is specified 
by the trajectory $B(L_i)$, along which the compression perpendicular to the magnetic field proceeds. 
As will be explained below, in lattice regularization it is natural to keep the flux fixed, and thus, 
the lattice measurements correspond directly 
to the $\Phi$-scheme. 
However, this does not represent a limitation of the lattice approach, since one can easily 
translate from one scheme into another. 
The pressure components for a general $B(L_i)$ trajectory (``general scheme'') can be found by 
combining our results for the longitudinal pressure and for the magnetization (both are contained in the ancillary files submitted to the arXiv),
\be
p_x^{\rm (general)} = p_z + \M \cdot L_x\frac{\partial (eB)}{\partial L_x}.
\label{eq:generalscheme}
\ee
This relation reproduces the $B$- and $\Phi$-schemes, Eq.~(\ref{eq:BPHI}), for the trajectories 
$B(L_i)=B$ and $eB(L_i)=\Phi/(L_xL_y)$, respectively.

Another important observable for the EoS is the interaction measure (trace anomaly),
\be
I\equiv\epsilon-p_x-p_y-p_z,
\label{eq:Idef}
\ee
which contains the energy of the medium and the three pressures. 
Thus, $I$ also depends on the scheme\footnote{One may understand the scheme-dependence of $I$ as follows. The trace anomaly represents the 
response to a rescaling of the length scale $\xi$ in the system. 
To define this rescaling unambiguously, the trajectory $B(\xi)$ has to be specified, 
i.e.\ a scheme has to be chosen. Hence, $I$ becomes scheme-dependent. 
As a simple example, consider the magnetic field in the 
absence of particles. Taking into account the energy $B^2/2$ of the magnetic field, one obtains 
$I^{(\Phi)}=0$, while $I^{(B)}=2B^2$. Notice that in the $\Phi$-scheme a dimensionless number characterizes 
the magnetic field, whereas in the $B$-scheme we introduced a dimensionful parameter 
into the system. This 
is reflected by the vanishing of the trace anomaly in the former case, and the nonzero value of $I$ in 
the latter.
}:
\be
I^{(B)} = \epsilon - 3p_z, \quad\quad\quad 
I^{(\Phi)} = \epsilon - 3p_z + 2\eBM,
\label{eq:Idef2}
\ee
whereas the energy density (like $p_z$) is by construction scheme-independent,
\be
\epsilon = I^{(B)} + 3p_z = I^{(\Phi)} + 3p_z -2\eBM.
\ee
Eqs.~(\ref{eq:fundeq}),~(\ref{eq:pzdef}) and~(\ref{eq:Idef2}) reveal that the 
entropy density can also be calculated as
\be
s = \frac{\epsilon + p_z}{T}.
\ee
Finally, the derivative of the magnetization with respect to $B$ at vanishing magnetic 
field gives the magnetic susceptibility,
\be
\chi_B = \left.\fracp{\M}{(eB)}\right|_{B=0} = -\frac{1}{V} \left.\frac{\partial^2 \F}{\partial (eB)^2}\right|_{B=0}.
\label{eq:defsusc}
\ee

\section{Lattice observables and methods}
\label{sec:latobs}

In what follows we consider a spatially symmetric lattice with isotropic lattice spacing $a$. Here the temperature and the three-volume are given by
\be
T=(N_t a)^{-1},\quad\quad\quad V=(N_s a)^3,
\label{eq:TandV}
\ee
where $N_s$ and $N_t$ are the number of lattice sites along the spatial and temporal directions, respectively.

Using conventional Monte-Carlo methods the free energy $\F=-T\log\Z$ 
itself is not accessible on the lattice, 
but only its derivatives with respect to the parameters of the theory. For the case of $2+1$ flavor QCD coupled to a constant external magnetic field, these parameters are the inverse gauge coupling $\beta=6/g^2$, the lattice quark masses $m_fa$ ($f=u,d,s$ labeling the flavors) and 
the magnetic flux $\Phi=(N_s a)^2eB$. 
In particular, in the staggered formulation of lattice QCD, $\Z$ is written as
\be
\Z = \int \D U e^{-\beta S_g} \prod_{f=u,d,s} \left[ \det M(U,a^2q_fB,m_fa)\right]^{1/4},
\label{eq:partfunc}
\ee
where $M=(\slashed{D}+m_f)a$ is the fermion matrix, and the quark charges are set to 
$q_d=q_s=-q_u/2=-e/3$. 
Note that the magnetic field has no dynamics, therefore the charge $q_f$ (or, the elementary charge $e$) and the magnetic field $B$ do not appear separately in the partition function, but always in the combination $q_fB$ (or $eB$). The constant Maxwell term $B^2/2$ is 
independent of the physical properties of the thermal QCD medium and plays no 
role in the thermodynamics of the system. It only enters in the renormalization 
prescription, see Sec.~\ref{sec:renorm} below. 

We work with the tree-level improved Symanzik gauge action $S_g$, and stout improved 
staggered quarks in the fermionic sector. The detailed simulation setup is described in Refs.~\cite{Aoki:2005vt,Bali:2011qj}. 
The quark masses are set to their physical values 
along the line of constant physics (LCP). 
This means that $m_fa$ are tuned as functions of $\beta$ in a way that ``physics remains the same'', that is to say, ratios of hadron masses measured on the lattice coincide with their 
experimental values. This defines the physical quark masses $m_f^{\rm ph}a$ for each value of $\beta$. 
In particular, our LCP is set by fixing the ratio of the kaon decay 
constant to the pion mass $f_K/M_\pi$ and the kaon decay constant to the kaon mass $f_K/M_K$. 
This results in the fixed ratio of quark masses $m_u=m_d\equiv m_{ud} = m_s/28.15$. 
The lattice spacing $a(\beta)$ is set using $f_K$. For additional 
details on this procedure, see Ref.~\cite{Borsanyi:2010cj}.

The derivatives of $\log\Z$ with respect to $\beta$ and $m_fa$ are the gauge action density and the quark condensate densities,
\be
a^4s_g=-\frac{1}{N_s^3N_t}\fracp{\log\Z}{\beta},\quad\quad\quad
a^3\bar\psi_f\psi_f=\frac{1}{N_s^3N_t}\fracp{\log\Z}{(m_fa)}.
\label{eq:densities}
\ee
The interaction measure, Eq.~(\ref{eq:Idef}), can be given in terms of the 
response of the free energy to an overall change of length scales in the 
system. On the lattice this amounts to a derivative with respect to the 
lattice spacing $a$. Employing the $a$-dependence of the lattice parameters 
$\beta$ and $m_fa$, the densities of Eq.~(\ref{eq:densities}) enter the 
$\Phi$-scheme interaction measure in the following way:
\be
I^{(\Phi)} = -\frac{T}{V} \left.\fracp{\log\Z}{\log a}\right|_{\Phi} = 
\fracp{\beta}{\log a} \,s_g - \sum_f \fracp{\log (m_f^{\rm ph}a)}{\log a} \,m_f\bar\psi_f\psi_f.
\ee
Note that for the $B$-scheme interaction measure (see Eq.~(\ref{eq:Idef2})), 
an additional term containing the derivative with respect to the lattice flux $a^2eB$ 
appears.

For convenience, we also define the change due to $B$ for any observable $X$ as
\be
\Delta X \equiv \left.X\right|_{B} - \left.X\right|_{0}.
\label{eq:deltadef}
\ee
The renormalization of the above observables will be discussed in Sec.~\ref{sec:renorm}.

\subsection{Flux quantization and methods to determine the magnetization}
\label{sec:fluxq}

Due to the periodic boundary conditions, the magnetic flux traversing the finite lattice is quantized as
\be
\Phi = (N_sa)^2 \cdot eB= 6\pi N_b,\quad\quad\quad N_b\in \mathds{Z},\quad0\le N_b<N_s^2,
\label{eq:quant}
\ee
where we took into account that the smallest charge in the system (that of the down quark) is $q_d=e/3$. 
Note that since the flux is quantized, the lattice setup automatically corresponds 
to the $\Phi$-scheme defined in Sec.~\ref{sec:thermo}.
Moreover, due to the quantization condition, differentiation with respect to $eB$ is 
in principle ill-defined and therefore the magnetization of Eq.~(\ref{eq:diffrels}) is not accessible directly. 
Recently, several methods were developed to circumvent this problem, which we summarize briefly 
below.
\begin{itemize}
 \item {\bf Anisotropy method.} 
One can make use of the relation~(\ref{eq:BPHI}) for the $\Phi$-scheme, and express the magnetization as the difference 
between the longitudinal and transverse lattice pressures. These can be measured as derivatives of $\log\Z$ with respect to anisotropy parameters. This approach was developed and successfully applied in Refs.~\cite{Bali:2013esa,Bali:2013owa}. The advantage of the method is that $\M$ is directly 
obtained as an expectation value for any $B$, 
while its drawback is that anisotropy renormalization 
coefficients also need to be determined.

\item {\bf Half-half method.} 
Instead of the uniform (and, thus, quantized) magnetic field, one can work with an inhomogeneous 
field which has zero flux, e.g. one that is positive in one half and negative in the other half 
of the lattice. Since the field strength is now a continuous variable, derivatives of $\log\Z$ 
with respect to $eB$ are well defined and can be measured on a $B=0$ lattice 
ensemble~\cite{Levkova:2013qda}. 
The second-order derivative directly gives the magnetic susceptibility. However, higher-order terms become increasingly noisy, which limits the applicability of the 
approach to low fields.
Note moreover that the discontinuities in the magnetic field may enhance finite volume effects. 

\item {\bf Finite difference method.} 
The derivative of $\log\Z$ with respect to $eB$ is an unphysical quantity 
due to the quantization Eq.~(\ref{eq:quant}). Still, this derivative 
can be measured for any real value of $N_b$, and 
its integral over $N_b$ between two integer values gives the change in $\log\Z$ between these two 
fluxes. In this way, $\log\Z(N_b+1)-\log\Z(N_b)$ is constructed as the integral of an oscillatory 
function. The method is in principle applicable for any magnetic field, but 
10-20 independent simulations are necessary to go from one integer flux to the next, making large magnetic fields computationally expensive~\cite{Bonati:2013lca,Bonati:2013vba}.

\item {\bf This work: generalized integral method.} 
The method we will use in the present paper is based on two observations: that magnetic fields 
have no effect in pure gauge theory, and that the infinite quark mass limit of QCD (at a fixed magnetic 
field $qB\ll m^2$) is pure gauge theory. 
Based on this, the change in $\log\Z$ due to the magnetic field can be expressed as an integral 
of the quark condensate differences $\Delta \bar\psi_f\psi_f$ over the quark masses, including unphysically 
heavy quarks. On a finite lattice, this 
integral is well regulated and can be calculated in a controlled manner by using 10-20 independent 
simulations for any given value of the magnetic field. Most of these simulations 
are at large quark masses, where the computation is significantly cheaper. 
Furthermore, the method automatically gives information on the mass-dependence of $\log\Z$ as well. 
This approach was sketched in Ref.~\cite{Bali:2013txa} and will be 
described in detail below.

\end{itemize}

\subsection{Renormalization}
\label{sec:renorm}

The free energy density contains additive divergences in the cutoff -- i.e.\ in the 
inverse lattice spacing. These divergences are independent of $eB$, except for one 
logarithmic divergence of the form $-\QEDb (eB)^2\log (\mu a)$, where $\mu$ is a 
renormalization scale. This term is canceled through a 
redefinition of the energy $B^2/2$ of the magnetic field itself~\cite{Schwinger:1951nm},
\be
\frac{B^2}{2} = \frac{B_r^2}{2} + \QEDb (eB)^2 \log (\mu a). 
\label{eq:Bdiv}
\ee
Eq.~(\ref{eq:Bdiv}) 
is equivalent to a simultaneous renormalization of the wave function (magnetic field $B$) and 
of the electric charge $e$. The combination $eB$ is renormalization group invariant and, as such, unaffected by this transformation:
\be
Z_e=1+2\,\QEDb e_r^2 \log(\mu a), \quad\quad
B^{2} = Z_e B_r^2, \quad\quad e^{2} = Z_e^{-1} e_r^2, \quad\quad eB=e_rB_r.
\ee
The purely magnetic contribution $B_r^2/2$ is trivial and can be omitted  
from the Lagrangian. Therefore, 
the renormalization of the free energy amounts to adding the counter-term $\QEDb (eB)^2 \log(\mu a)$
to $\Delta f$. 
In the following it will be advantageous to consider the extensive quantity $\Delta\log\Z=-L^4 \Delta f$ 
at zero temperature, in a box of four-volume $L^4$. The counter-term then takes the form 
$-\QEDb \Phi^2 \log(\mu a)$ with the flux $\Phi=L^2 eB$.
The coefficient $\QEDb$ of the divergence is related to the QED $\beta$-function~\cite{Abbott:1981ke,Elmfors:1993bm,Dunne:2004nc}. Since the 
magnetic field is external, i.e.\ there are no $\mathrm{U}(1)$ degrees of freedom in the system, 
only the lowest order QED $\beta$-function coefficient $\QEDb$ appears in $Z_e$ 
(however, with a full dependence on the QCD coupling, see Eq.~(\ref{eq:betacorr}) below).

\subsubsection{Charge renormalization -- free case}
\label{sec:chrenfree}

It is instructive to first discuss the renormalization procedure in the free case -- 
i.e.\ for electrically charged quarks in the absence of strong interactions. 
In this case the free energy can be calculated analytically (see, e.g., 
Refs.~\cite{Elmfors:1993bm,Dunne:2004nc,Endrodi:2013cs}). 
We consider quark flavors of charges $q_f$ and, for simplicity, we assume 
degenerate masses $m_f=m$ for all $f$. The discussion is easily generalized to unequal masses. For $N_c=3$ colors, 
the QED $\beta$-function coefficient reads
\be
\QEDb^{\rm free} = \sum_f b_{1f}^{\rm free}, \quad\quad\quad b_{1f}^{\rm free} = \frac{N_c}{12\pi^2} \cdot (q_f/e)^2.
\label{eq:betafree}
\ee
At zero temperature, the expansion of $\Delta \log\Z$ in the magnetic field is given by
\be
\Delta \log\Z^{\rm free}_r = \QEDb^{\rm free} \cdot \Phi^2 \cdot \log (m_f a) + \mathcal{O}(\Phi^4)
- \QEDb^{\rm free} \cdot \Phi^2 \cdot \log(\mu a),
\label{eq:T0fBfree}
\ee
where we also included the counter-term. 
Taking the derivative with respect to the mass of the quark flavor $f$, we obtain the corresponding quark condensate 
at $T=0$,
\be
\Delta \bar\psi_f\psi_f^{\rm free}= 
\frac{1}{L^4}\fracp{\Delta \log\Z^{\rm free}_r}{m_f} = 
b_{1f}^{\rm free} \frac{(eB)^2}{m_f}+ \mathcal{O}((eB)^4),
\label{eq:pbpfree}
\ee
showing that the condensate contains no $B$-dependent divergences, and that it 
is also independent of the renormalization scale $\mu$.
Note also that the sign of the magnetic field-induced change in the condensate 
is, to leading order, determined by the sign of $b_{1f}^{\rm free}$. Since QED is not 
asymptotically free, $b_{1f}^{\rm free}$ is positive and 
the condensate undergoes magnetic catalysis at $T=0$ to quadratic 
order in $eB$ (we have already presented this argument in Refs.~\cite{Endrodi:2013cs,Bali:2013txa}).
Note that approaching the chiral limit (i.e. $eB/m^2\to\infty$), the magnetic 
field-expansion in 
Eq.~(\ref{eq:pbpfree}) becomes 
ill-defined. In fact, in the $m_f\to0$ limit $\Delta \bar\psi_f\psi_f^{\rm free}$ 
vanishes (see, e.g., Ref.~\cite{Shovkovy:2012zn}) for any magnetic field. 
However, the condensate difference is expected 
to be positive if any weak attractive interaction is turned on~\cite{Gusynin:1995nb}.

Let us now calculate the interaction measure. We resort to the $\Phi$-scheme of Sec.~\ref{sec:thermo}, 
as this is the natural one in the lattice setup. 
Contrary to the case of the condensate, 
here the counter-term also contributes a finite term $\QEDb^{\rm free} (eB)^2$. 
The remainder of Eq.~(\ref{eq:T0fBfree}) depends only on the 
combination $m_fa$, thus the derivative with respect to $\log a$ is equivalent to that with respect to $\log m_f$. Using Eqs.~(\ref{eq:T0fBfree}) and~(\ref{eq:pbpfree}), we therefore obtain
\be
\Delta I^{{\rm free}(\Phi)}_r=-\frac{1}{L^4}\cdot\left.\fracp{\log\Z^{\rm free}_r}{\log a}\right|_{\Phi} = 
-\sum_f m_f\Delta \bar\psi_f\psi_f^{\rm free} + \QEDb^{\rm free} (eB)^2
=
\mathcal{O}((eB)^4),
\label{eq:Ifree}
\ee
which is again finite and $\mu$-independent. 
The trace anomaly difference contains the two well-known sources of scale violation~\cite{Adler:1976zt}: 
the classical 
breaking through the condensates\footnote{Note that the usual 
definition of the condensate (with $\bar\psi_f\psi_f<0$) differs from our convention by a minus sign.} and the anomalous one through the running of the electric 
charge. In our case, the two contributions cancel each other to 
$\mathcal{O}((eB)^2)$, since at this order $a$ drops out of Eq.~(\ref{eq:T0fBfree}). 
We shall return to this observation below. 

We have seen that the condensate difference and the trace anomaly difference 
are independent of the 
renormalization scale. However, in order to define the renormalized 
free energy and pressures, we need to specify $\mu$ in Eq.~(\ref{eq:T0fBfree}). 
Setting the renormalization scale equal to the mass, $\mu=m_f$, means that 
all terms quadratic in the magnetic field are canceled. 
This scheme\footnote{
We remark that renormalization schemes with different choices for the scale $\mu$ have also been 
used in the literature. For example, $\mu$ is taken to be proportional to $\sqrt{eB}$ in the schemes 
employed in Refs.~\cite{Menezes:2008qt,Fraga:2012fs}, which are connected to our choice by a finite (albeit mass-dependent) renormalization. Our scheme has the advantage that the leading magnetic field-dependence 
of the total free energy is simply $B_r^2/2$. Moreover, the $m\to\infty$ limit of the magnetization 
vanishes, in accordance with the expectation that magnetic fields should have no effect on static non-relativistic 
particles (see the discussion in Ref.~\cite{Endrodi:2013cs}).
} is intrinsic to the Schwinger proper time 
representation~\cite{Schwinger:1951nm}, and coincides with the one used 
in Ref.~\cite{Endrodi:2013cs}.
Since in this scheme the expansion of $\log\Z$ starts as $(eB)^4$ at $T=0$, 
so do the expansions of the pressures, of the energy density and of $\eBM$. 
Therefore $I$ -- being a linear combination of the former -- 
has an expansion starting with a quartic term as well, consistent with Eq.~(\ref{eq:Ifree}).

We remind the reader that we excluded the renormalized pure magnetic energy $B_r^2/2$ above, which 
depends explicitly on the renormalization scale $\mu$. To restore this term in $\log\Z$
one needs to add
\be
-\frac{B_r^2(\mu)}{2} = -\frac{(eB)^2}{2} \cdot \frac{1}{4\pi \alpha_{\rm em}(\mu)},
\ee
where $\alpha_{\rm em}(\mu)=e_r^2(\mu)/(4\pi)$ is the running QED coupling defined 
at the scale $\mu$.

\subsubsection{Charge renormalization -- full QCD}
\label{sec:chrenfullqcd}

We proceed by applying the renormalization prescription discussed for free quarks above to the case
of full QCD. With the strong interactions taken into account, $\QEDb$ will contain QCD 
corrections, which, in a perturbative expansion in the strong coupling $g$ take the form
\be
\QEDb(a) = \QEDb^{\rm free} \cdot \bigg[ 1 + \sum_{i\ge1} c_i \,g^{2i}(1/a) \bigg]
\xrightarrow{a\to0} \QEDb^{\rm free}, 
\label{eq:betacorr}
\ee
where the coefficients $c_i$ are 
independent of the quark masses and have been 
calculated in the $\overline{\rm MS}$ scheme up to $i=4$ in Ref.~\cite{Baikov:2012zm}. 
Note that the running of the QCD coupling -- governed by the QCD $\beta$-function -- 
induces a dependence of $\QEDb$ on the regulator, which on the lattice amounts to 
a dependence on the lattice spacing $a$. 
Thus, due to the asymptotically free nature of the strong interactions, 
QCD corrections vanish in the continuum limit, and $\QEDb(a)$ approaches its free value, 
as indicated in Eq.~(\ref{eq:betacorr}). 
We will see that for the lattice spacings we employ, these corrections are already tiny, 
see the right panel of Fig.~\ref{fig:pertfit} below.

The consistency with charge renormalization ensures that the free energy is again of the 
form Eq.~(\ref{eq:T0fBfree}). Contrary to the free case, inside the logarithm of the divergent term, 
an additional dimensionful hadronic scale $\Lambda_{\rm H}$ appears, which may depend on $m_f$.  
The expansion of $\Delta \log\Z$ at zero temperature then reads
\be
\Delta \log\Z_r = \QEDb(a) \cdot \Phi^2 \cdot \log (\Lambda_{\rm H} a) + \mathcal{O}(\Phi^4) 
- \QEDb(a) \cdot \Phi^2 \cdot \log(\mu a).
\label{eq:T0fB}
\ee
From this -- in analogy to the free case -- we can extract the leading dependence of the condensate difference
and of the interaction measure difference on the magnetic field:
\be
\Delta \bar\psi_f\psi_f = 
\QEDb(a) \cdot \frac{(eB)^2}{\Lambda_{\rm H}} \cdot\fracp{\Lambda_{\rm H}}{m_f} + \mathcal{O}((eB)^4), \quad\quad\quad
\Delta I^{(\Phi)}_r=
\mathcal{O}((eB)^4).
\label{eq:pbpfull}
\ee
We again conclude that the $B$-dependent divergence is absent from the 
condensate~\cite{Bali:2011qj,Bali:2013esa}. Moreover, in the 
renormalization group invariant combination $m_f\Delta \bar\psi_f\psi_f$, multiplicative 
divergences cancel as well.
To quadratic order in $eB$ the sign of the change of the condensate is 
related to the sign of $\QEDb$ and to that of $\partial \Lambda_{\rm H}/\partial m_f$, 
which we will revisit in Sec.~\ref{sec:condbeta}.

The interaction measure difference is also explicitly finite, as noted in 
Ref.~\cite{Bali:2013esa}, where we determined the gluonic and fermionic contributions 
to $\Delta I^{(\Phi)}$ separately. Similarly to the free case, Eq.~(\ref{eq:Ifree}), 
$\Delta I^{(\Phi)}_r$ receives a 
finite contribution from the counter-term in $\Delta \log\Z_r$,
\be
\frac{1}{L^4}\frac{\partial}{\partial \log a} \left[ \QEDb(a) \Phi^2 \log(\mu a) \right] = (eB)^2\cdot \left[ \QEDb(a) - \log(\mu a) \cdot\fracp{\QEDb}{g^2}\cdot\fracp{g^2}{\log (1/a)} \right]
\xrightarrow{a\to0} (eB)^2 \cdot \QEDb^{\rm free},
\ee
which, due to Eq.~(\ref{eq:betacorr}), equals its free-case equivalent in the continuum limit 
(the QCD $\beta$-function damps the second term in the square brackets as $a\to0$). 
In the continuum limit, the contribution from the counter-term results in  
a cancellation to $\mathcal{O}((eB)^2)$ in the total interaction 
measure, as already stated in Eq.~(\ref{eq:pbpfull}). For later reference, the renormalization 
of $\Delta I^{(\Phi)}$ thus reads
\be
\Delta I^{(\Phi)}_r = \Delta I^{(\Phi)} + \QEDb^{\rm free} \cdot (eB)^2.
\label{eq:Irdef}
\ee

To discuss the renormalization of $\log\Z$ itself, we have to specify the 
renormalization scale.
We may again choose $\mu$ such that the 
quadratic term in $\log\Z$ at $T=0$ is completely subtracted in the renormalization process: $\mu=\Lambda_{\rm H}$.
This is the equivalent of the on-shell renormalization scheme in the free case. 
The renormalization prescription 
for the free energy (and, similarly, for the longitudinal pressure) at $T=0$ in 
this scheme reads
\be
f_r = (1-\mathcal{P}) [f], \quad\quad\quad
p_{z,r} = (1-\mathcal{P}) [p_z],
\label{eq:frdef}
\ee
where we defined $\mathcal{P}$ as the operator that projects out the $\mathcal{O}((eB)^2)$ term from an observable $X$:
\be
\mathcal{P}[X] = (eB)^2 \cdot \lim_{eB\to0} \left.\frac{X}{(eB)^2}\right|_{T=0}.
\label{eq:Projdef}
\ee
We remark that at finite temperature, thermal contributions induce additional finite 
terms that are quadratic in $eB$. Thus, the subtraction of $\P[X]$ is to be
performed at $T=0$, as indicated in Eq.~(\ref{eq:Projdef}).

\subsection{The integral method at nonzero magnetic fields}
\label{sec:intmethod}

To determine the free energy -- or, equivalently, the longitudinal pressure $p_z$, see Eq.~(\ref{eq:pzdef}) 
-- on the lattice, we employ a variation of the so-called integral method~\cite{Engels:1990vr}. The 
basic idea is to construct $p_z$ by integrating its partial derivatives in Eq.~(\ref{eq:densities}) along a 
particular path in the parameter space spanned by the parameters $\{\beta,m_fa,\Phi\}$. 
Since the magnetization is not accessible as a derivative (see Sec.~\ref{sec:fluxq}), one is only allowed to integrate along a constant-$\Phi$ trajectory in this parameter space. For a lattice of fixed size $N_s^3\times N_t$, the magnetic field thus changes as $eB\sim a^{-2}\sim T^2$ along such a path.

Specifically, we consider a trajectory at constant $\Phi$, from $\beta_1$ to $\beta_2$ 
with the quark masses tuned along the LCP $m_f^{\rm ph}a$. Then, 
the integral method is written down for the change $\Delta p_z$ in the pressure: 
the difference of $\Delta p_z$ 
at the two endpoints equals the integral of the gradient of $\Delta p_z$ along this 
trajectory. Using the definitions Eq.~(\ref{eq:densities}) of the subtracted 
lattice observables $\Delta s_g$ and $\Delta \bar\psi_f\psi_f$, we obtain
\be
\frac{\Delta p_z(\Phi,T_2;\beta_2)}{T_2^4} - \frac{\Delta p_z(\Phi,T_1;\beta_1)}{T_1^4}
 = N_t^4 \int_{\beta_1}^{\beta_2} \dd\beta \left[  -a^4\Delta s_g + \sum_f \fracp{(m_f^{\rm ph}a)}{\beta} \cdot a^3\Delta \bar\psi_f\psi_f \right].
\label{eq:pres1}
\ee
Here, the endpoints $\beta_i$ of the integral correspond to the temperatures $T_i$, and 
tuning the quark masses along the LCP resulted in the factor $\partial(m_f^{\rm ph}a)/\partial \beta$. 

The expression~(\ref{eq:pres1}) gives the difference between the dimensionless pressure differences 
$\Delta p_z/T^4$ at two distinct temperatures for a given $\Phi$. To determine the change in the pressure at one 
temperature, for each such $\Phi$ additional information is necessary, which corresponds to fixing an 
integration constant. 
In the conventional integral method~\cite{Engels:1990vr} at $B=0$, one exploits the fact that $p/T^4$ vanishes 
at zero temperature, therefore the integration constant at $T=0$ is zero. Here, this method is not applicable, since in the presence of a magnetic field, the zero-temperature pressure is no longer zero, 
see Eq.~(\ref{eq:T0fB}). Instead, we propose to use a different region of the parameter space to fix the integration constant, namely the $m_fa=\infty$ line, which corresponds to pure gauge theory plus 
free static quarks. Since the external magnetic field couples only to quarks, in pure gauge theory $B$ has by definition no effect, and $\Delta p_z(\Phi,T)$ is 
given solely by the contribution $\Delta p_z^{\rm free}(\Phi,T)$ of 
free heavy quarks, which is naively expected to vanish for any $T$ and any finite $\Phi$ 
in the limit $m_f^2\gg qB$. 

\begin{figure}[ht!]
 \centering
 \includegraphics[width=7.8cm]{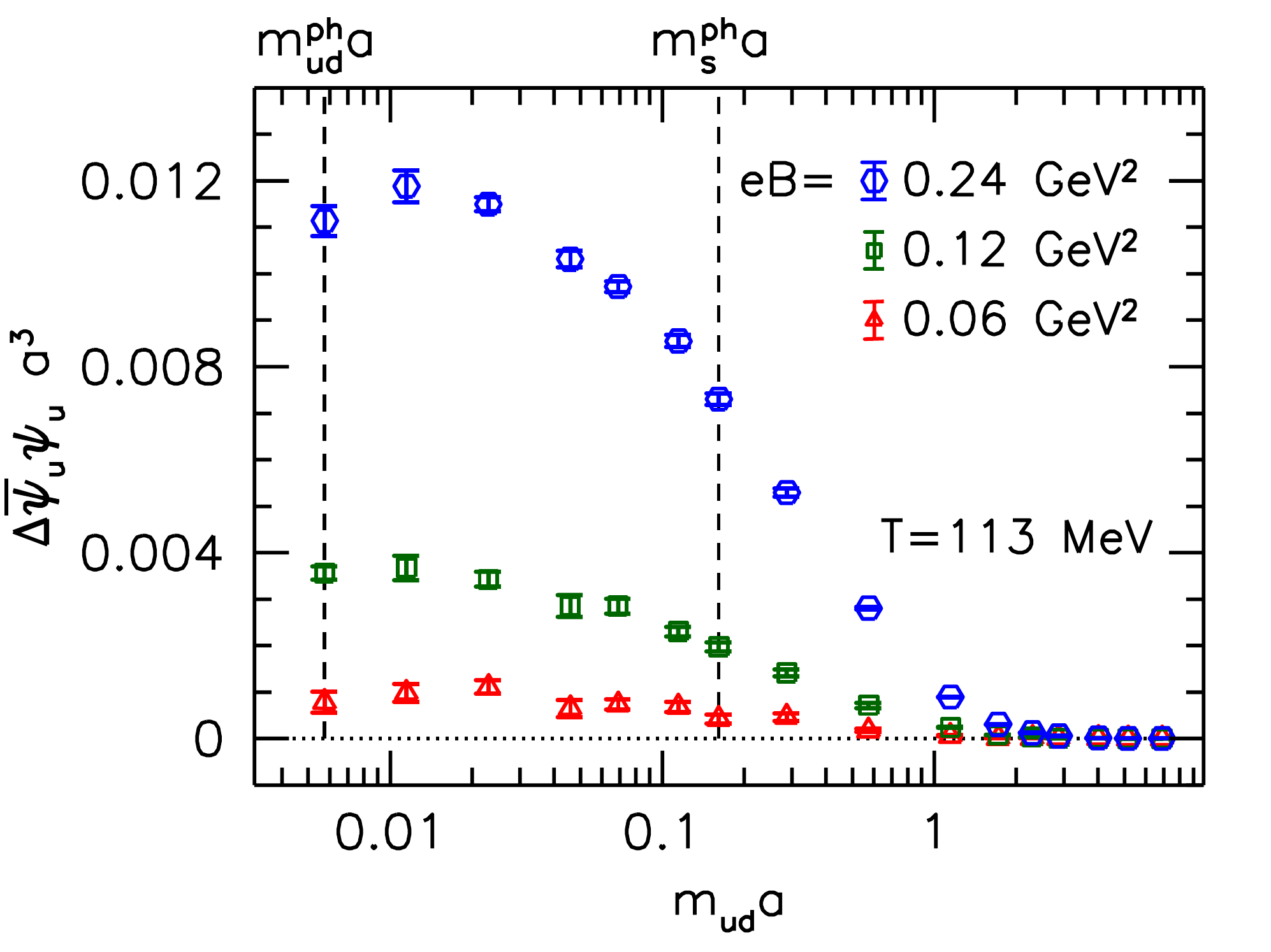} \quad\quad
 \includegraphics[width=7.8cm]{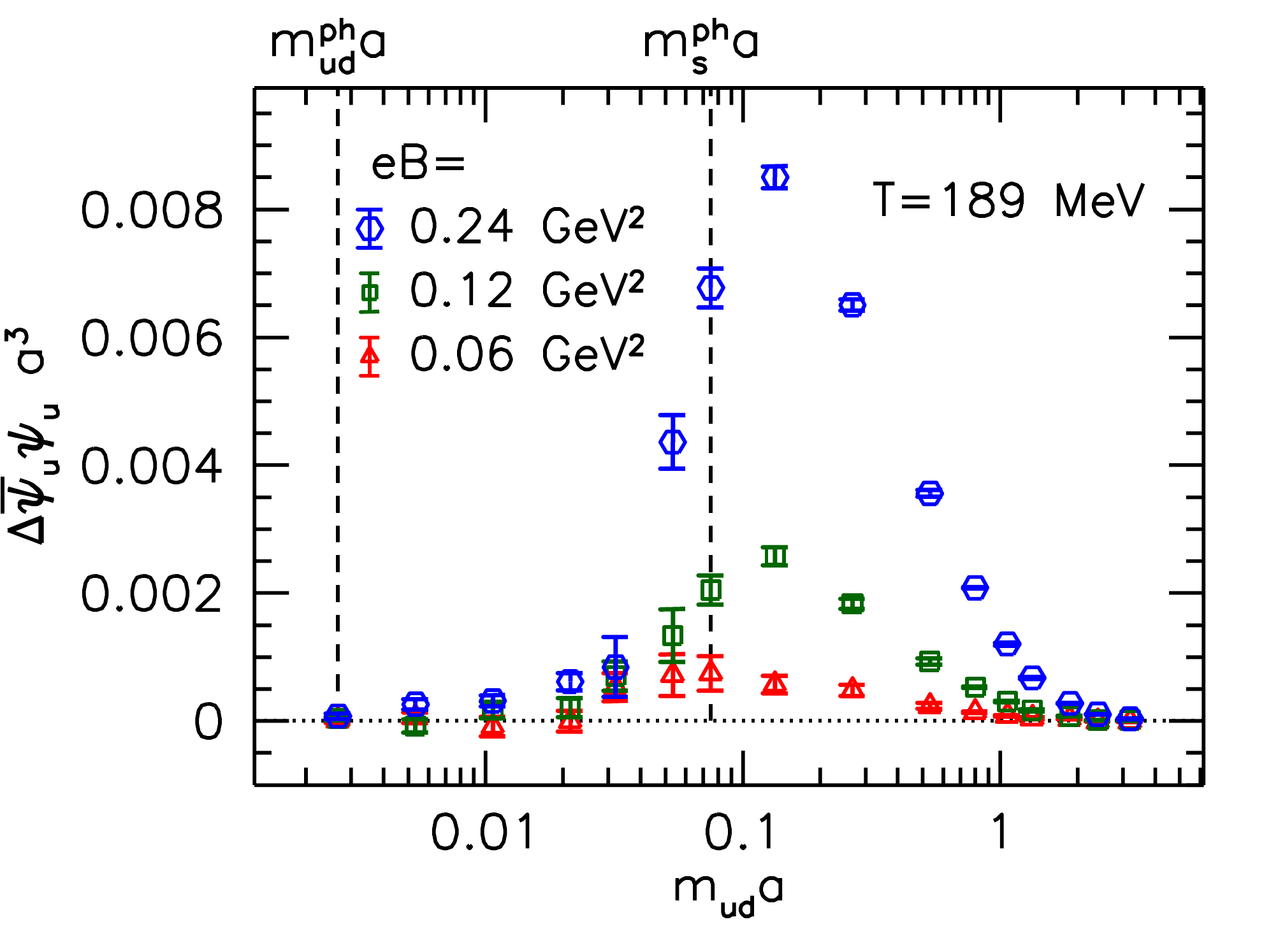}
\vspace*{-.3cm}
 \caption{ \label{fig:deltapbp}
The change of the condensate $\Delta \bar\psi_u\psi_u$ in lattice units, as a function of the quark mass 
on the $N_t=6$ lattices at $T=113\textmd{ MeV}$ (left panel) and at $T=189\textmd{ MeV}$ (right panel). 
Different colors encode different magnetic fields. The dashed lines indicate the physical light and strange quark masses.
 }
\end{figure}

However, in the continuum theory the {\it bare} $\Delta p_z^{\rm free}$ for static quarks contains the 
ultraviolet divergent term $\propto \QEDb^{\rm free} \Phi^2$ (higher orders 
in $\Phi$ vanish in the static limit), see Eq.~(\ref{eq:T0fBfree}). Therefore $\Delta p_z^{\rm free}$ 
only vanishes in the infinite mass limit after the renormalization has been carried out. 
Nevertheless, in the lattice 
regularization $\Delta p_z^{\rm free}$ is suppressed as $1/(m_fa)^4$ once the quark mass exceeds the lattice 
scale $1/a$, see App.~\ref{app:expand} and the discussion in Sec.~\ref{sec:condbeta} below. 
Thus we conclude that 
at finite lattice spacings, $\Delta p_z$ vanishes in the asymptotic quark mass limit. Therefore, 
integrating down to the physical quark masses $m_f^{\rm ph}a$ at fixed $\beta$, we 
obtain for an arbitrary temperature
\be
\frac{\Delta p_z(\Phi,T;\beta)}{T^4}
= -N_t^4 \sum_f \int_{m_f^{\rm ph}a}^{\infty} \dd (m_fa) \,a^3\Delta \bar\psi_f\psi_f.
\label{eq:pres2}
\ee
Thus, the pressure difference is expressed as an integral of $\Delta \bar\psi_f\psi_f$ over all 
higher-than-physical quark masses. 
In practice we first integrate over the two light quark masses up to the point where all three masses 
coincide (the $N_f=3$ theory with different quark charges). Second we integrate over the quark masses simultaneously\footnote{Note that any integration path in the \{$m_ua,m_da,m_sa\}$ space between the physical point and $\{\infty,\infty,\infty\}$ is admissible and gives the same result.} up to $m_fa=\infty$.
The integrand for the up quark is shown in Fig.~\ref{fig:deltapbp} as a function of the light lattice quark mass for three values of the magnetic field, as measured on the $N_t=6$ lattices. At $T=113 \textmd{ MeV}$ (left panel of the figure), the difference $\Delta \bar\psi_u\psi_u$ is positive, reflecting the well-known magnetic catalysis of the 
condensate at low temperatures, see, e.g., Refs.~\cite{Gusynin:1995nb,Bali:2012zg}. As the mass is increased and the quark decouples, this difference eventually approaches zero.

In the right panel of Fig.~\ref{fig:deltapbp} the same observable is shown, but in the high-temperature phase, at $T=189 \textmd{ MeV}$. At the physical light quark mass the difference is close to zero (see also the results presented in Ref.~\cite{Bali:2012zg}), and as the mass is increased a peak-like structure is revealed. 
This structure is a consequence of the strong dependence of the transition temperature $T_c$ 
on the light quark mass: Around $T_c$ chiral symmetry is restored, the 
condensate is strongly suppressed and magnetic catalysis is not effective anymore. 
While at the physical point $T_c\approx 150 \textmd{ MeV}$~\cite{Borsanyi:2010bp,Bazavov:2011nk}, 
in pure gauge 
theory $T_c\approx 260 \textmd{ MeV}$~\cite{Boyd:1996bx}. 
At $T=189 \textmd{ MeV}$ we start in the chirally restored phase, but as the masses are increased, 
at some point the transition line is crossed and we enter 
the chirally broken phase where magnetic catalysis is dominant and $\Delta \bar\psi_u\psi_u$ is large. 
Eventually, for $m\to\infty$ the difference again approaches zero.
We note that the down quark condensate shows a very similar behavior for both temperatures 
$T=113 \textmd{ MeV}$ and $T=189\textmd{ MeV}$.

Through Eq.~(\ref{eq:pres2}) we have determined the integration constant.
Now, complemented by Eq.~(\ref{eq:pres1}), the change of the pressure $\Delta p_z(\Phi,T)$ 
can be determined at any temperature $T$ and at any magnetic flux $\Phi$. 
Next, the renormalization is 
performed according to Eq.~(\ref{eq:frdef}) to obtain the renormalized pressure 
$\Delta p_{z,r}(\Phi,T)$. 
The resulting curves are interpolated to compute $\Delta p_{z,r}(eB,T)$ 
for any $T$ and $eB$. This 
is finally shifted by the zero-field pressure, which we take from Ref.~\cite{Borsanyi:2013bia}, 
to obtain the full pressure for a range of magnetic fields and temperatures. 
From the longitudinal pressure, all other thermodynamic observables can be calculated using the 
relations of Sec.~\ref{sec:thermo}.

\subsection{Lattice ensembles}

Before presenting the results, we briefly describe the lattice ensembles we used. 
These consist of two sets of lattice configurations: one at high temperatures, necessary 
for the determination of the $T$-dependence of the EoS, and one at effectively zero temperature, necessary 
for the renormalization. 
The high-$T$ ensemble contains $N_t=6$, $8$ and $10$ lattices with various values of the 
inverse gauge coupling $\beta$, such that the temperature range 
$113\textmd{ MeV}< T < 300 \textmd{ MeV}$ can be scanned and a continuum estimate 
can be given (note that at a fixed temperature, the lattice spacing is proportional to $1/N_t$ such 
that the continuum limit corresponds to $N_t\to\infty$, see Eq.~(\ref{eq:TandV})). 
These configurations correspond to physical quark masses, tuned along the LCP as discussed at the 
beginning of Sec.~\ref{sec:latobs}. 
This ensemble was mainly generated in Ref.~\cite{Bali:2011qj} for the study of the phase diagram and is supplemented in the present analysis by configurations at $T=250 \textmd{ MeV}$ and $T=300\textmd{ MeV}$. 

Based on detailed comparisons to our zero-temperature $24^3\times 32$ ensembles 
(see Sec.~\ref{sec:quadratic}), it turned out that the renormalization factors can 
be determined reliably at our lowest `finite-temperature' point, $T=113 \textmd{ MeV}$.
At this temperature we included two additional lattice spacings with $N_t=12$ and $16$, allowing 
for a determination of the renormalization factors down to small lattice spacings, and a matching 
with perturbation theory. For each $N_t$ we generated configurations ranging from $m_{ud}=m_{ud}^{\rm ph}$ up 
to $m_{ud}=1200 \cdot m_{ud}^{\rm ph}$. 
The simulation parameters are listed in Table~\ref{tab:lats}.

\begin{table}[ht!]
\setlength{\tabcolsep}{3pt}
\setlength{\extrarowheight}{4pt}
\fontsize{10.}{10.5}\selectfont
 \centering
 \begin{tabular}{|c|c||c|c|c|c|c|c|}
\hline
   & \raisebox{1pt}{$m_{ud}/m_{ud}^{\rm ph}$} & $24^3\times6$ & $24^3\times8$ & $28^3\times10$ & $36^3\times12$ & $48^3\times16$ & $24^3\times32$ \\ \hline
  low-$T$ & $1\ldots1200$ & $\beta=3.45$ & $\beta=3.55$ & $\beta=3.625$ & $\beta=3.695$ & $\beta=3.81$ & $\beta=3.45, 3.55$ \\ \hline
  high-$T$ & 1 & $\beta=3.45\ldots3.81$ & $\beta=3.55\ldots 3.94$ & $\beta=3.625\ldots 4.06$ &  &  & \\ \hline
 \end{tabular}
 \caption{\label{tab:lats}
 Summary of our lattice ensembles.}
\end{table}

\section{Results}
\label{sec:results}

\subsection{Condensates, the $\mathbf{\beta}$-function and a comment on magnetic catalysis}
\label{sec:condbeta}

\begin{figure}[ht!]
 \centering
 \includegraphics[width=7.8cm]{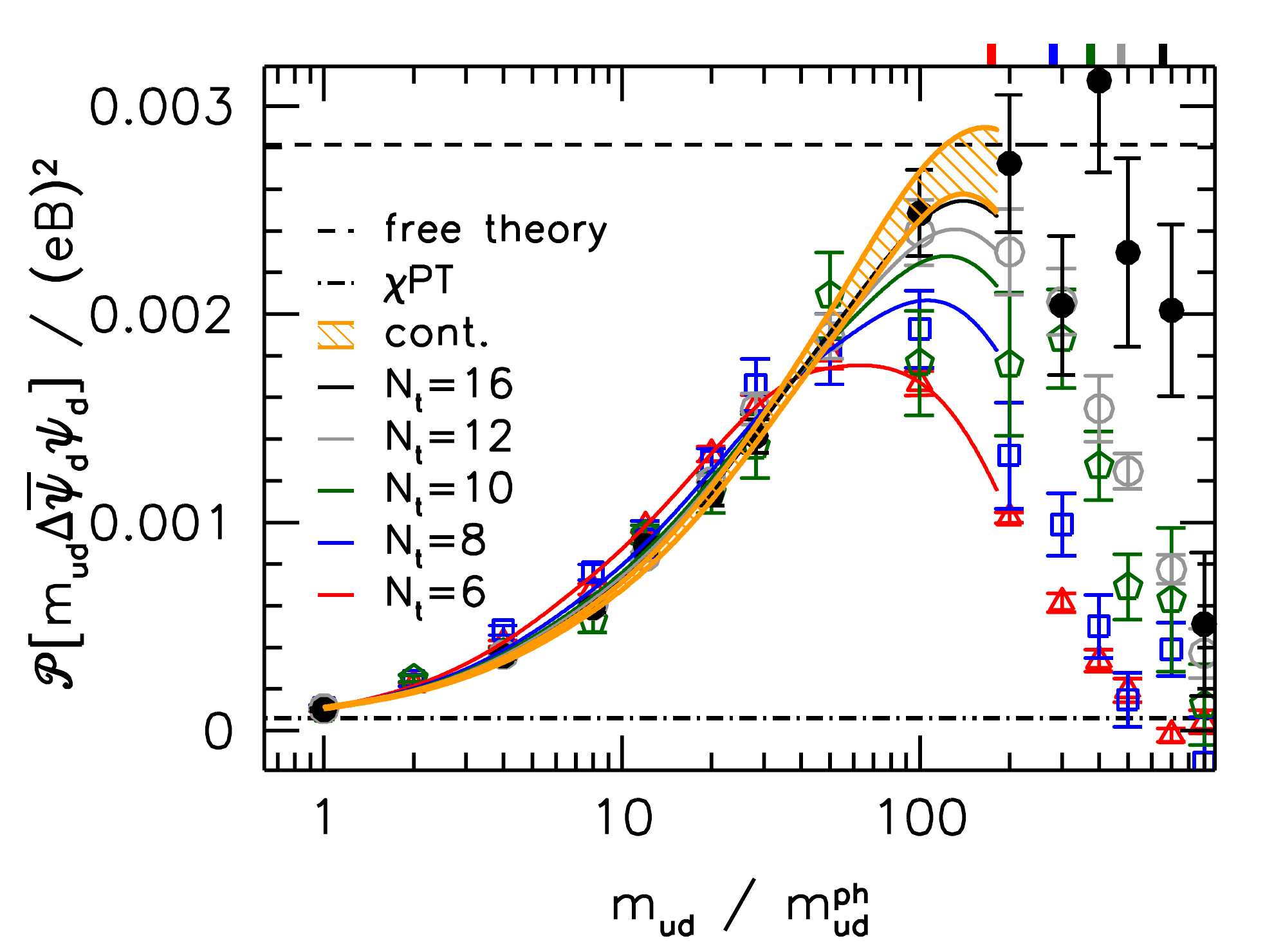}
\vspace*{-.3cm}
 \caption{ \label{fig:mpbp}
The $\mathcal{O}((eB)^2)$ contribution to the integrand of Eq.~(\protect\ref{eq:pres2}) 
normalized by $(eB)^2$
for five different lattice spacings and the continuum 
extrapolation. 
The free theory prediction (dashed line) and the expectation from $\chi$PT (dashed-dotted line) 
are also shown. For asymptotically large quark masses ($m_{ud}a=1$ for each lattice 
spacing is marked by the colored bars in the upper right corner) 
lattice artefacts become large and $\P[m_{ud}\Delta \bar\psi_d\psi_d]$ drops. }
\end{figure}

We start the presentation of our results with additional details on and implications of 
the generalized integral method. 
Let us consider the integrand on the right hand side of Eq.~(\ref{eq:pres2}), and expand 
it in powers of $eB$. 
For asymptotically large quark masses, quarks and gluons decouple, and 
$\Delta\bar\psi_f\psi_f$ approaches its free theory value. 
In this limit, we obtain from Eq.~(\ref{eq:pbpfree}),
\be
\frac{\P [m_f\cdot \Delta \bar\psi_f\psi_f^{\rm free}] }{(eB)^2}= 
b_{1f}^{\rm free},
\label{eq:freepredpbp}
\ee
where $\P$ is the projector defined in Eq.~(\ref{eq:Projdef}). 
In Fig.~\ref{fig:mpbp} we plot $\P[m_{ud}\Delta\bar\psi_d\psi_d]$ normalized by $(eB)^2$ 
as a function of the light quark mass at $T=113\textmd{ MeV}$. We perform a combined continuum extrapolation of 
all five lattice spacings ($N_t=6,8,10,12$ and $16$), 
assuming $\mathcal{O}(a^2)$ discretization errors. 
The plotted combination has 
a finite continuum limit (see the discussion after Eq.~(\ref{eq:pbpfull})), 
but discretization errors become large when the lattice quark mass $m_{ud}a$ approaches unity. 
For asymptotically large masses the change of the condensate is proportional to $(m_{ud}a)^{-5}$, 
as shown in App.~\ref{app:expand}. 
The finer the lattice, the later this lattice artefact sets in, and the larger are the 
quark masses that can be reached. 
With our present lattice spacings we can control
the continuum extrapolation up to $m_{ud}/m_{ud}^{\rm ph}\approx 100-200$. Here the extrapolated 
values are consistent with the free theory prediction, Eq.~(\ref{eq:freepredpbp}),
showing that in the continuum limit $\Delta\bar\psi_f\psi_f\propto 1/m_f$ for large masses $m_f\gg \Lambda_{\rm QCD}$, to quadratic order in $eB$. 
This implies that the 
$\mathcal{O}((eB)^2)$ contribution to the integral on the 
right hand side of Eq.~(\ref{eq:pres2}) diverges logarithmically. 
On the lattice this divergence is regulated 
by the inverse lattice spacing such that the cutoff $m_fa\approx1$ plays the role of the upper limit 
of the mass-integral, see the sharp drop in Fig.~\ref{fig:mpbp} for asymptotically large masses
$m_f\gtrsim a^{-1}$. 
In the continuum limit the logarithmic divergence reappears (cf.\
Eq.~(\ref{eq:T0fB})), and has to be subtracted via charge renormalization.

Let us now determine the chiral limit of the combination shown in Fig.~\ref{fig:mpbp} 
using chiral perturbation theory ($\chi$PT). The lowest excitation in this case is the charged pion, 
implying that to leading order $\Lambda_{\rm H} = m_\pi$ in Eqs.~(\ref{eq:T0fB}) and~(\ref{eq:pbpfull}). Moreover, 
due to the spin-zero nature of the pion, 
the scalar QED $\beta$-function appears instead of the spinor $\beta$-function. 
Using the Gell-Mann-Oakes-Renner 
relation $m_\pi^2F^2=\bar\psi\psi(0) (m_u+m_d)$ we obtain for the light flavors $f=u,d$,
\be
\frac{m_f}{m_\pi^2}\cdot\fracp{m_\pi^2}{m_f} = \frac{1}{2}, \quad\quad\quad\quad\quad\quad
\frac{\P [m_f\cdot \Delta \bar\psi_f\psi_f^{\chi \rm PT}] }{(eB)^2}= 
\frac{b_{1f}^{\rm free,scalar}}{4N_c} = \frac{b_{1f}^{\rm free}}{16N_c},
\label{eq:chiPTpredpbp}
\ee
as was already pointed out within the Hadron Resonance Gas (HRG) model~\cite{Endrodi:2013cs}. 
To derive Eq.~(\ref{eq:chiPTpredpbp}), we considered 
equal masses for the light flavors $m_u=m_d$. 
Note that the first relation in Eq.~(\ref{eq:chiPTpredpbp}) is understood to hold at $B=0$, where 
the charged and neutral pion masses are equal. 
We indicate the $\chi$PT 
prediction in Fig.~\ref{fig:mpbp} by the dashed-dotted line, showing a good agreement with 
the lattice data at physical quark masses, see also the comparison in Ref.~\cite{Bali:2012zg}.

Altogether we observe that the QCD quark condensate, as determined in a fully non-perturbative treatment, interpolates between the $\chi$PT prediction at small masses and the free-theory 
limit at large masses. 
We have seen in Eq.~(\ref{eq:pbpfull}) that the sign of the condensate $\Delta \bar\psi_f\psi_f$ equals  
the product of the sign of the factor $\partial \Lambda_{\rm H}/\partial m_f$ 
and that of $\QEDb$. While for large quark masses $m_f\gg \Lambda_{\rm QCD}$ one has 
$\Lambda_{\rm H}=m_f$, towards 
the chiral limit $\Lambda_{\rm H}=m_\pi$ such that the first factor is in both 
cases positive. 
It would be quite unexpected to have an intermediate mass where this factor 
turned negative, nevertheless, we did not find any strict proof of this positivity. 
In any case we conclude that the $\mathcal{O}((eB)^2)$ magnetic catalysis of the quark 
condensate and the positivity of the scalar/spinor QED $\beta$-function are 
intimately related phenomena. This picture was first described in Ref.~\cite{Endrodi:2013cs} and 
also discussed in Ref.~\cite{Bali:2013txa}. 
Note that the above argument concerns the $\mathcal{O}((eB)^2)$ behavior of 
the condensate and does not address what happens in the large-$B$ limit, where 
the dimensional reduction is expected to be the dominant driving force of magnetic 
catalysis~\cite{Gusynin:1995nb}. 

\subsection{Quadratic contribution to the EoS}
\label{sec:quadratic}

\begin{figure}[b]
\centering
\vspace*{-.3cm}
\includegraphics[width=7.8cm]{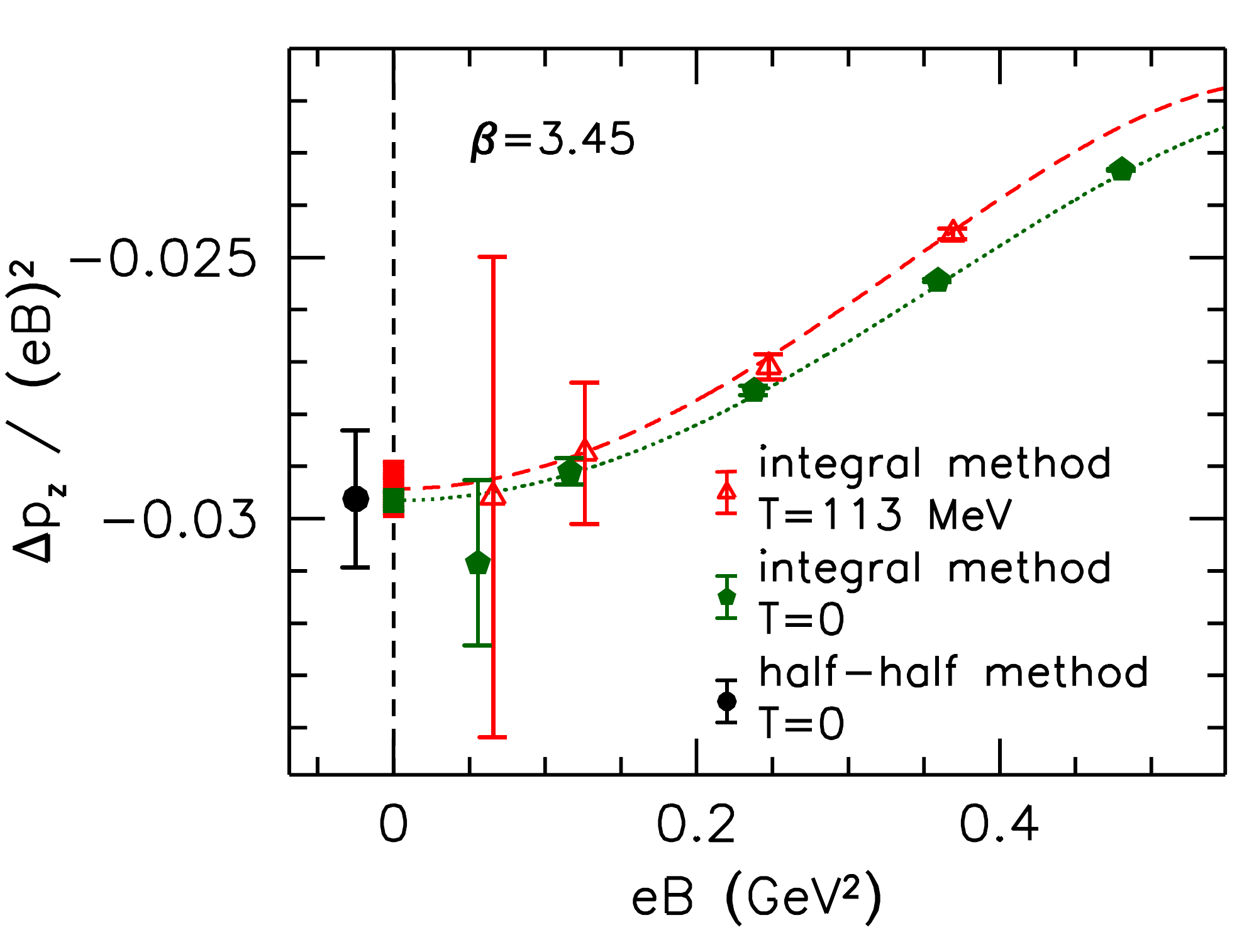} \quad\quad
\includegraphics[width=7.8cm]{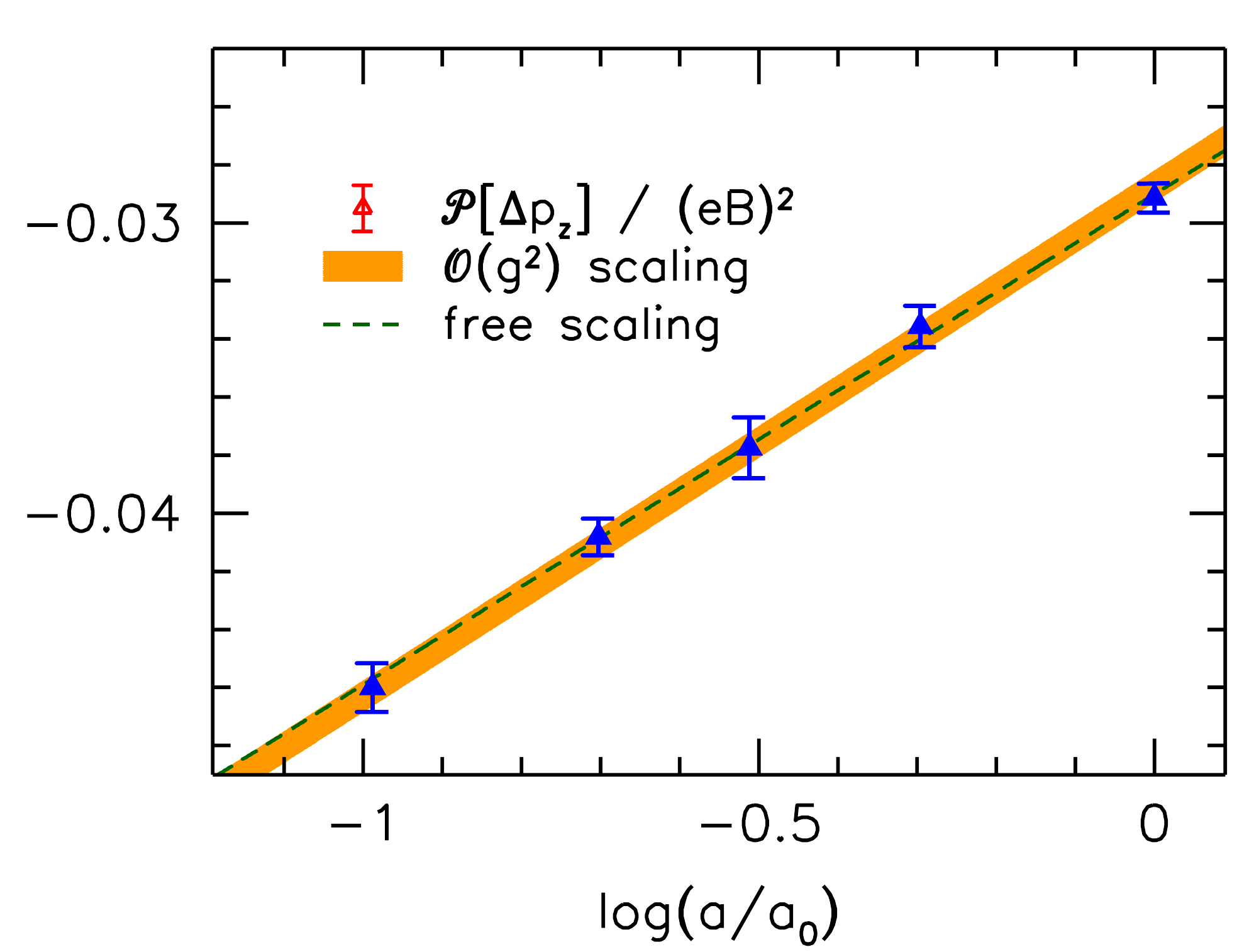}
\vspace*{-.2cm}
\caption{\label{fig:pertfit}
Left panel: magnetic field-dependence of the bare longitudinal pressure difference at low temperatures 
at one lattice spacing. 
The results of the integral method at $T=113 \textmd{ MeV}$ and at $T=0$, 
and the result of the half-half method at $T=0$ are compared (the points are slightly 
shifted horizontally for better visibility).  
Right panel: quadratic contribution to the bare longitudinal pressure at $T=113\textmd{ MeV}$ against 
the lattice spacing in units of $a_0=1.47 \textmd{ GeV}^{-1}$. 
}
\end{figure}

Next we calculate the $\mathcal{O}((eB)^2)$ contribution to the longitudinal pressure at 
effectively zero temperature. This term will be subtracted through charge renormalization, according 
to Eq.~(\ref{eq:frdef}). 
We perform the integral Eq.~(\ref{eq:pres2}) to determine $\Delta p_z$ at various values of the magnetic flux (see the left panel of Fig.~\ref{fig:pertfit}). To extract the $\mathcal{O}((eB)^2)$ contribution, we fit the data to a quadratic function in $(eB)^2$ and take the $B\to0$ limit of 
$\Delta p_z / (eB)^2$, represented by the bars at $eB=0$ in the figure. 
We perform this analysis for the $24^3\times6$ ensemble 
at our lowest finite-temperature point, $T=113\textmd{ MeV}$, 
and on the $24^3\times32$ ensemble, which corresponds to $T=0$.
The $B\to0$ limits at these two temperatures are found to coincide within our 
statistical errors. This implies 
that thermal contributions to the $\mathcal{O}((eB)^2)$ pressure are still 
strongly suppressed at $T=113\textmd{ MeV}$, in agreement with our previous 
findings from the anisotropy method~\cite{Bali:2013owa}, with the results of 
Ref.~\cite{Levkova:2013qda} using the half-half method and with those of 
Ref.~\cite{Bonati:2013vba} using the finite difference method (for a brief 
description of these approaches, see Sec.~\ref{sec:fluxq}). 
Therefore, we conclude that within our present statistics, 
it is safe to use $T=113 \textmd{ MeV}$ as the reference 
temperature for the quadratic subtraction. 
This observation allows us to omit expensive lattice simulations at $T=0$, 
and to substitute these by cheaper runs at a finite but low temperature.

\begin{figure}[b]
\centering
\vspace*{-.3cm}
\includegraphics[width=7.8cm]{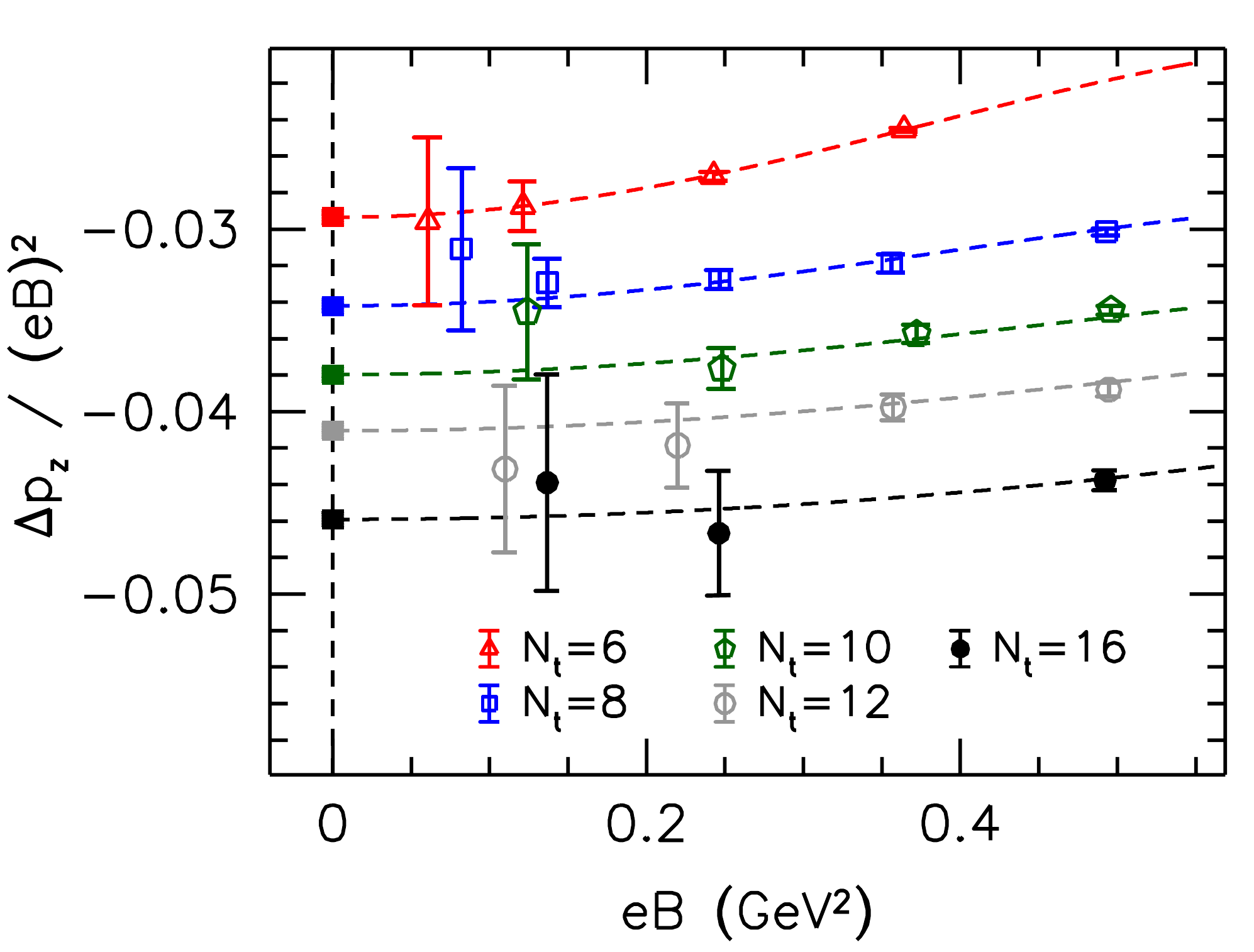} \quad\quad
\includegraphics[width=7.8cm]{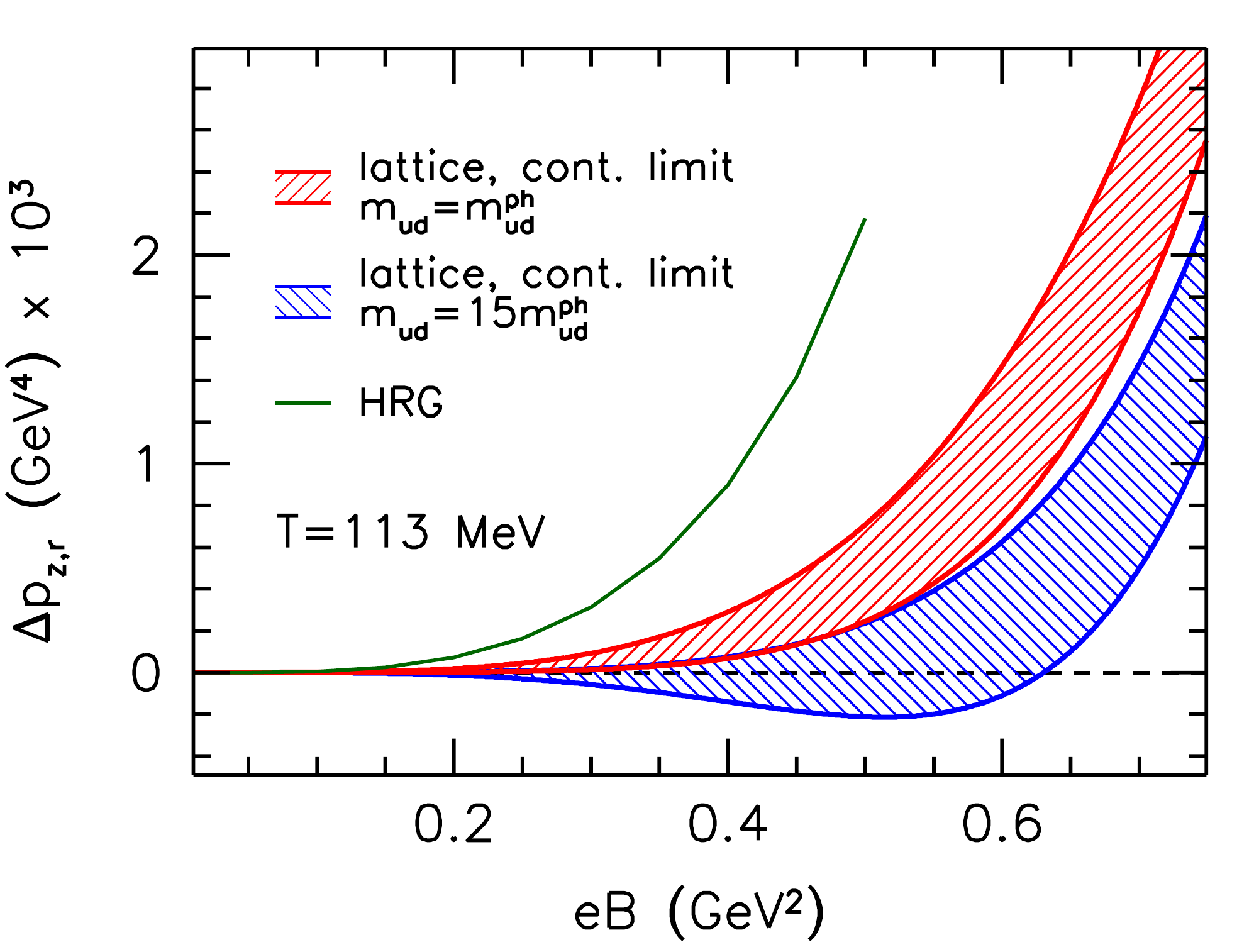}
\vspace*{-.3cm}
\caption{\label{fig:pzsubtr}
Left panel: combined extrapolation in the magnetic field and interpolation 
in the lattice spacing according to \protect Eq.~(\ref{eq:fitfunc}) using 
several values of the magnetic flux and five lattice spacings. Only data points with $eB<0.5\textmd{ GeV}^2$ 
are shown. 
Right panel: continuum limit of the renormalized magnetization for physical $m_{ud}=m_{ud}^{\rm ph}$ 
(red dashed band) and heavier-than-physical $m_{ud}=15\cdot m_{ud}^{\rm ph}$ quark masses (blue dashed band), and a comparison to the HRG model.
}
\vspace*{-.6cm}
\end{figure}

In addition, we also check the $eB\to0$ extrapolation of our results by directly 
determining the second derivative of $\Delta p_z$ with respect to the magnetic field at $eB=0$ using the 
half-half method at $T=0$. For this measurement we employ the setup of Ref.~\cite{Levkova:2013qda} and use $400$ noisy estimators to 
calculate the trace of the necessary operators (for details see Ref.~\cite{Levkova:2013qda}). The result is indicated by the black point in the left panel of Fig.~\ref{fig:pzsubtr}, showing that the two methods agree perfectly. 
Note that in the generalized integral method we use results at all $B>0$ to extract the 
quadratic part, resulting in smaller errors. Increasing instead the statistics at $B=0$ 
for the half-half method, we could also improve the signal-to-noise ratio of the latter, 
however finite volume effects should also be studied carefully in this case. 
Since the results at $eB>0$ will be in any case necessary 
for calculating the higher order contributions to the renormalized pressure, it is advantageous 
to use the generalized integral method to extract the quadratic term as well. 

We proceed by discussing the dependence of $\P[\Delta p_z]$ on $a$. 
First we perform the $B\to0$ extrapolation separately for each of our five lattice spacings. 
The resulting values are expected to lie on the curve $\QEDb(a)\cdot \log(\Lambda_{\rm H} a)$, see Eq.~(\ref{eq:T0fB}). 
We consider the universal one-loop QCD corrections to $\QEDb(a)$ -- i.e. terms up to $i=1$ in Eq.~(\ref{eq:betacorr}).
The strong coupling in the lattice scheme is defined as $g^2(1/a)=6/\beta(a)$. 
The so obtained function $\QEDb(a)\cdot \log(\Lambda_{\rm H}a)$ 
is fitted to the data with $\Lambda_{\rm H}$ considered as a free 
parameter. The result is indicated by the 
orange band in the right panel of Fig.~\ref{fig:pertfit}. 
For comparison we also carry out a similar fit in the free case, which corresponds to a simple 
linear fit with fixed slope $\QEDb^{\rm free}=0.0169$
(which is obtained using Eq.~(\ref{eq:betafree}) for three flavors). 
The two fits agree within errors for the 
whole range, indicating that for our lattice spacings, the QCD corrections to the 
QED $\beta$-function in Eq.~(\ref{eq:betacorr}) are smaller than our statistical errors.  

Exploiting the fact that the free scaling describes our data to a very good accuracy, 
we consider an alternative strategy to determine the quadratic contribution to $\Delta p_z$.
Namely, we fit the results for $\Delta p_z/(eB)^2$ at all magnetic fields and lattice spacings 
together using a combined inter/extrapolation in $a$ and $eB$. 
We consider the fit function 
\be
\frac{\Delta p_z}{(eB)^2} = c_0+b_1^{\rm free} \log\left(a/a_0\right) + (eB)^2\cdot \left(c_1+c'_1 a^2+c''_1 a^4\right) + (eB)^4 \cdot \left(c_2+c'_2 a^2\right),
\label{eq:fitfunc}
\ee
which takes into account the logarithmic divergence of the constant term 
(with the free scaling coefficient). Here $a_0=1.47 \textmd{ GeV}^{-1}$ is our largest lattice spacing (corresponding to $N_t=6$). 
We found it
necessary to include $\mathcal{O}(a^4)$ lattice discretization effects in the 
$(eB)^2$ part and $\mathcal{O}(a^2)$ terms in the $(eB)^4$ part of the fit function.
The results of this combined fit are shown in the left panel of Fig.~\ref{fig:pzsubtr}. 
Considering higher orders in the 
magnetic field or more lattice artefacts in Eq.~(\ref{eq:fitfunc}) 
did not improve the quality of the fit. 
The $eB\to0$ limits (colored bars in the figure) equal $\P[p_z]/(eB)^2$, giving the logarithmically
divergent term $c_0+b_1^{\rm free}\log(a/a_0)$ that is subtracted via charge renormalization, as a function of the 
lattice spacing. 
The fitted 
coefficients are listed in Table~\ref{tab:chren}. 
As a cross-check we carried out a similar fit with $b_1$ as a free parameter, resulting in 
a value consistent with the expected continuum value $b_1^{\rm free}$. 
Matching Eq.~(\ref{eq:fitfunc}) with Eq.~(\ref{eq:T0fB}) we read off that at the physical 
value of the quark masses
\be
\Lambda_{\rm H}(m_{ud}^{\rm ph})=e^{c_0/b_1^{\rm free}}/a_0=0.120(9) \textmd{ GeV}.
\ee
The scale $\Lambda_{\rm H}$ depends on the regularization scheme (i.e.\ on the lattice 
action). However, towards the chiral limit it is expected to approach a 
hadronic scale, $\Lambda_{\rm H}^{\rm \chi PT}=m_\pi$ (see the discussion in 
Sec.~\ref{sec:condbeta}), so that this scheme-dependence should only be mild. 
We stress that $\Lambda_{\rm H}$ is no free parameter but is automatically 
determined by the lattice implementation of the renormalization prescription
Eq.~(\ref{eq:frdef}). $\Lambda_{\rm H}$ will appear below in the perturbative 
description of the pressure as an input from the lattice side.

\begin{table}[ht!]
\setlength{\extrarowheight}{2pt}
 \centering
 \begin{tabular}{|c|c|c|c|}
   \hline
   $c_0$ & $c_1$ & $c'_1$ & $c''_1$ \\ \hline
   $-0.0294(5)$ & $0.006(6)\textmd{ GeV}^{-4}$ & $0.011(7)\textmd{ GeV}^{-2}$ & $0.003(2)$  \\ \hline \hline
   $c_2$ & $c'_2$ & $b_1^{\rm free}$ & $a_0$ \\ \hline
   $0.007(8) \textmd{ GeV}^{-8}$ & $-0.025(9) \textmd{ GeV}^{-6}$ & $0.0169$ & $1.47 \textmd{ GeV}^{-1}$ \\ \hline
 \end{tabular}
 \caption{\label{tab:chren}Parameters of the fit function Eq.~\protect(\ref{eq:fitfunc}).}
\end{table}

Subtracting terms quadratic in $eB$ gives the renormalized pressure 
$\Delta p_{z,r} = (1-\P)[\Delta p_z]$ at $T=113\textmd{ MeV}$. 
Its continuum limit is given by the $a\to0$ limit of our fit function Eq.~(\ref{eq:fitfunc}) 
and is shown in the right panel of Fig.~\ref{fig:pzsubtr}. 
Note that the positivity of $\Delta p_{z,r}$ indicates the response to be paramagnetic.
Besides 
the curve for the physical quark masses, we also include here the results obtained for heavier-than-physical quark masses. The analysis is the same in this case except for the lower endpoint of the 
integral in Eq.~(\ref{eq:pres2}) which we set to $m_{ud}=15\cdot m_{ud}^{\rm ph}$. These light 
quark masses correspond to a pion mass of about $500 \textmd{ MeV}$, and 
the fit according to Eq.~(\ref{eq:fitfunc}) gives $\Lambda_{\rm H}(15m_{ud}^{\rm ph}) = 159(10) \textmd{ MeV}$. 
The plot shows that the pressure 
is clearly less sensitive on $B$ as quarks become heavier. In addition we also indicate the 
Hadron Resonance Gas (HRG) model prediction~\cite{Endrodi:2013cs} for the physical pion mass. 
We will get back to the visible discrepancy between the lattice results and the model curve 
in Sec.~\ref{sec:renormEoS} below.

\subsection{Complete magnetic field dependence of the EoS}
\label{sec:renormEoS}

With the quadratic contribution determined as a function of the lattice spacing, we can 
carry out the renormalization of the pressure according to Eq.~(\ref{eq:frdef}), and 
that of the interaction measure using Eq.~(\ref{eq:Irdef}) for arbitrary 
temperatures. Using these two renormalized 
observables, all other EoS-related quantities can be calculated via the thermodynamic 
relations of Sec.~\ref{sec:thermo}. 

\begin{figure}[h]
\centering
\vspace*{-.2cm}
\includegraphics[width=7.8cm]{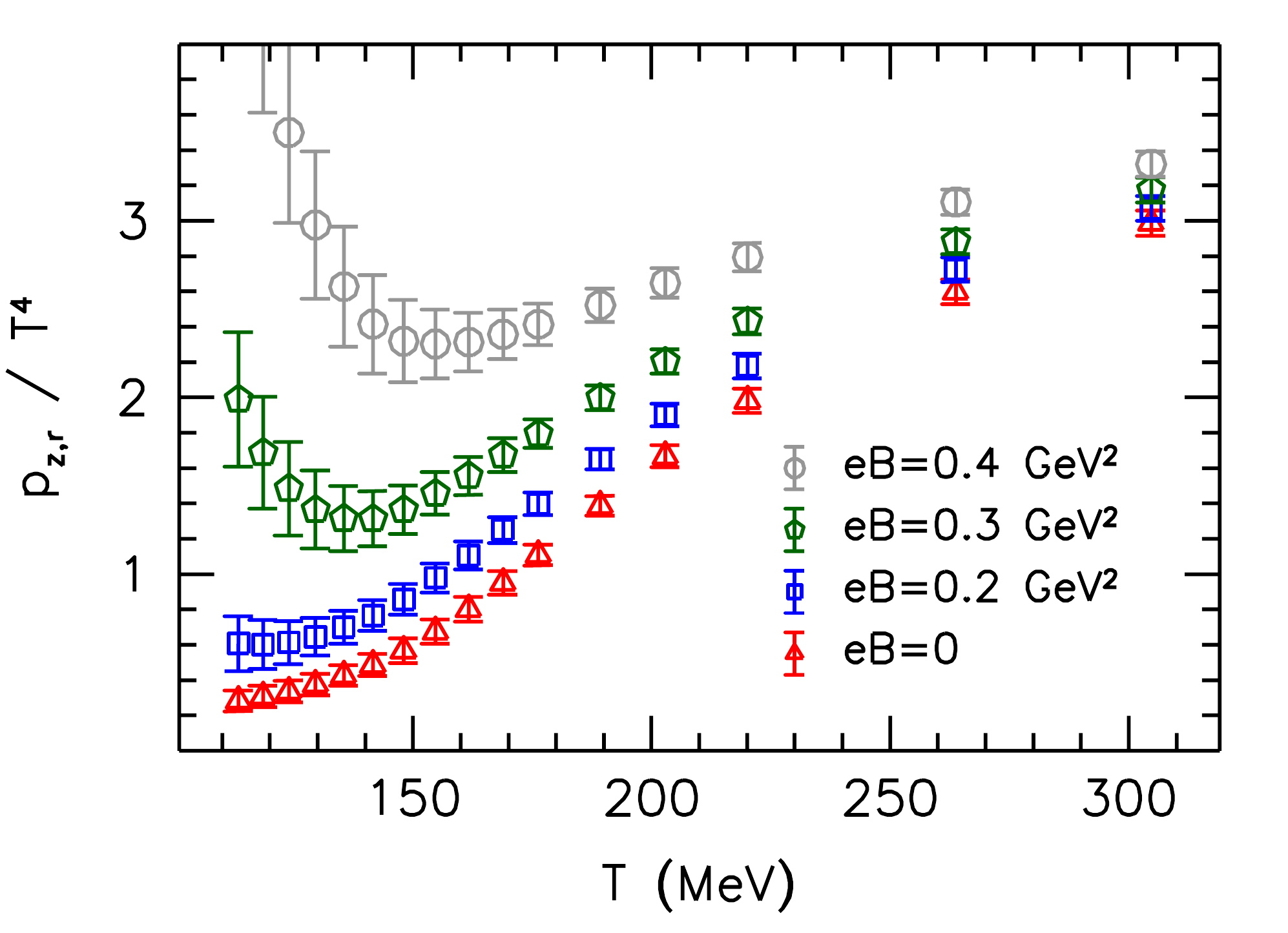} \quad\quad
\includegraphics[width=7.8cm]{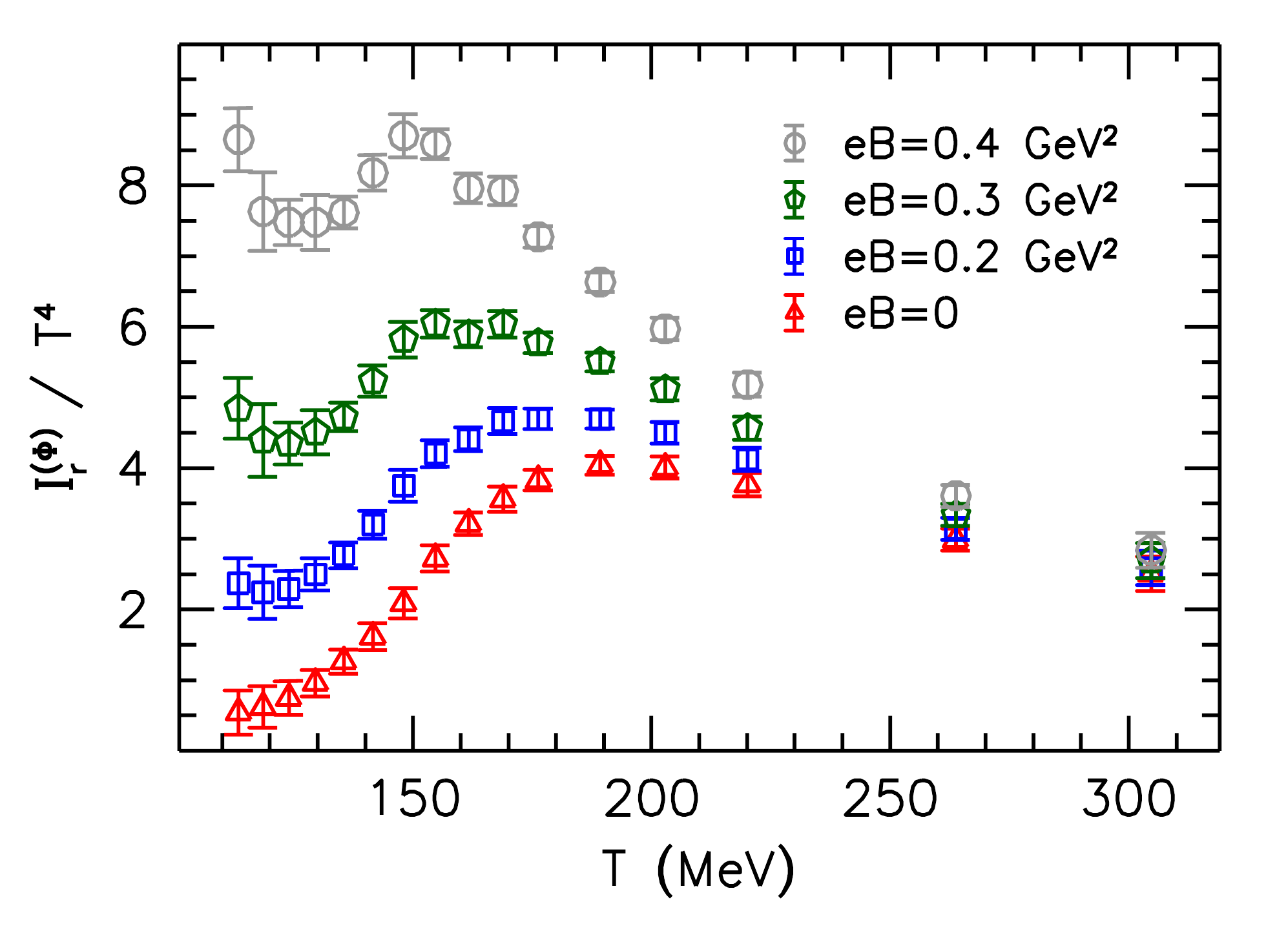}
\vspace*{-.4cm}
\caption{
\label{fig:eos_norm}
Longitudinal pressure (left panel) and interaction measure in the $\Phi$-scheme (right panel), 
normalized by $T^4$ as functions of the temperature, measured on our $N_t=6$ ensemble. 
Both $p_{z,r}$ and $I^{(\Phi)}_r$ are nonzero at $T=0$, which shows up as a quartic divergence at $T=0$. 
}
\end{figure}

\begin{figure}[t]
\centering
\vspace*{-.2cm}
\includegraphics[width=7.8cm]{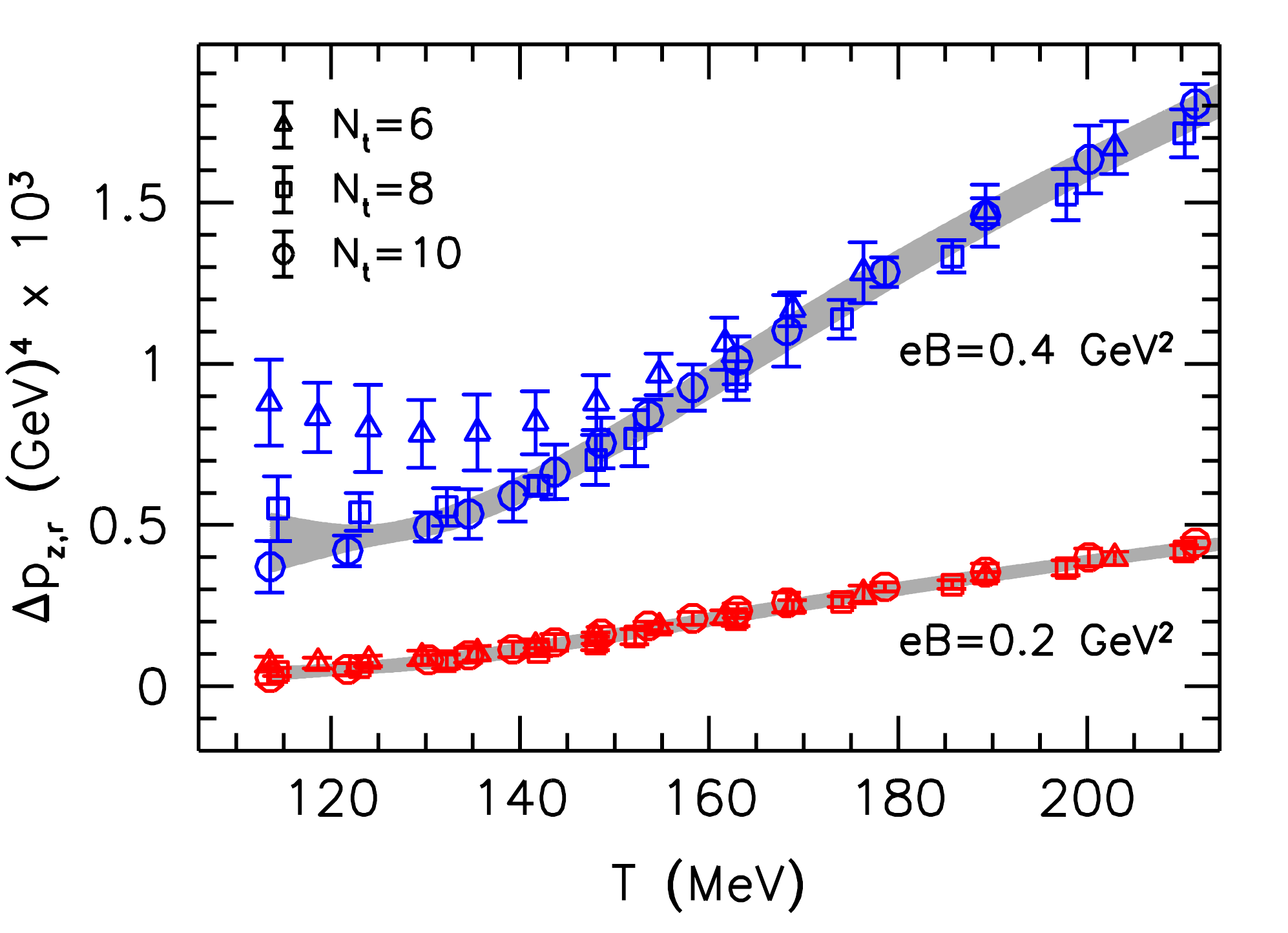} \quad\quad
\includegraphics[width=7.8cm]{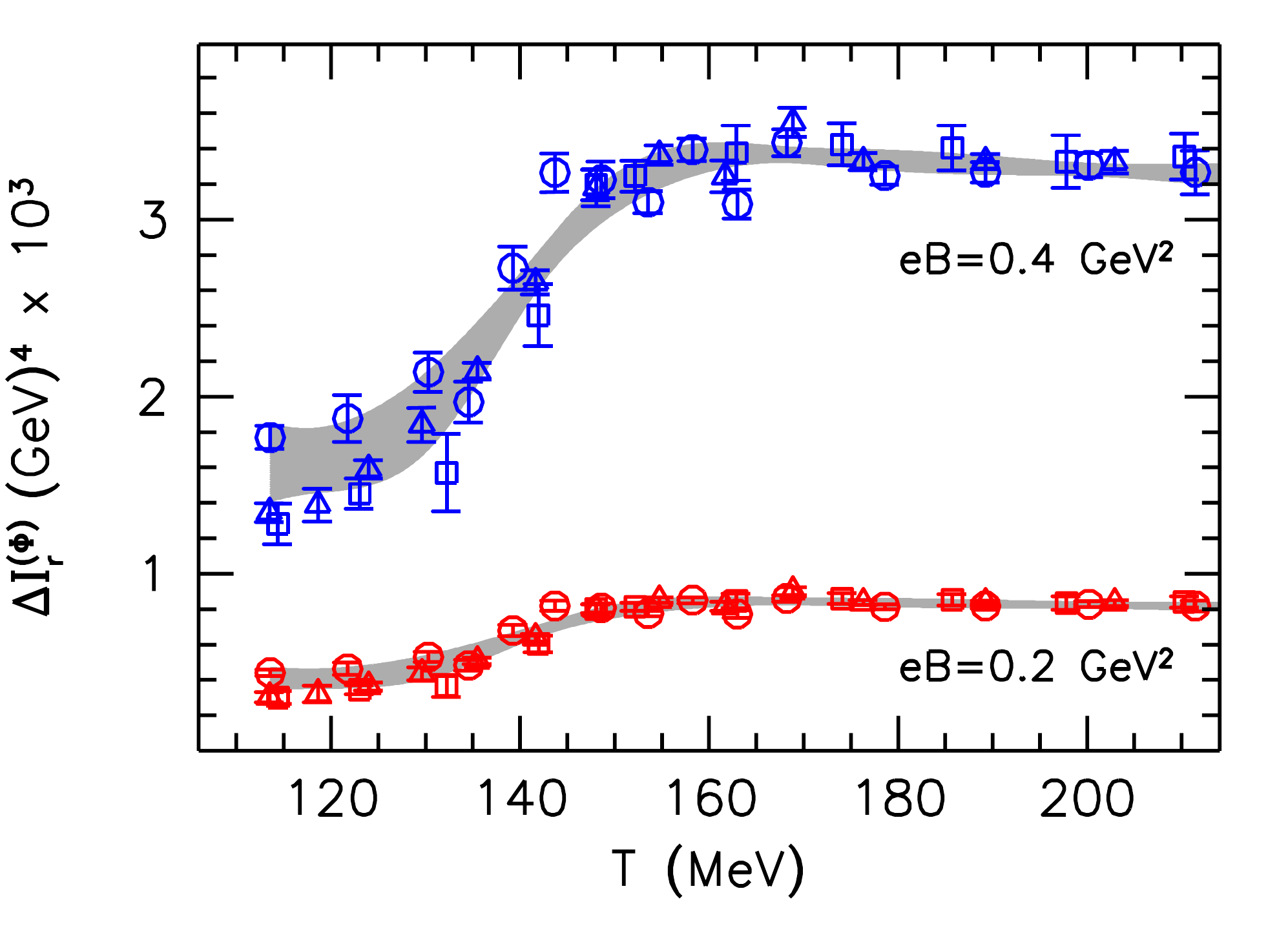}
\includegraphics[width=7.8cm]{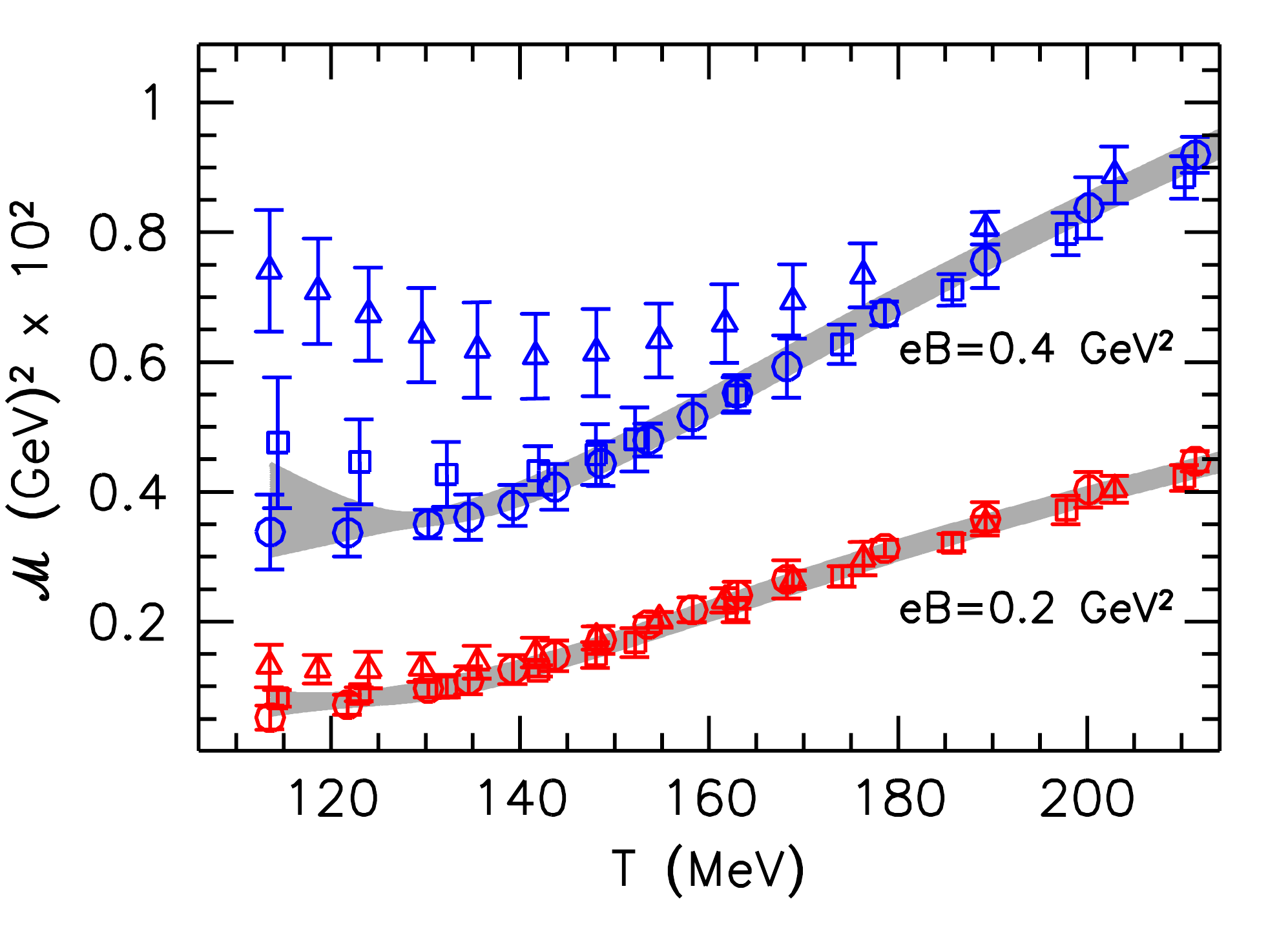} \quad\quad
\includegraphics[width=7.8cm]{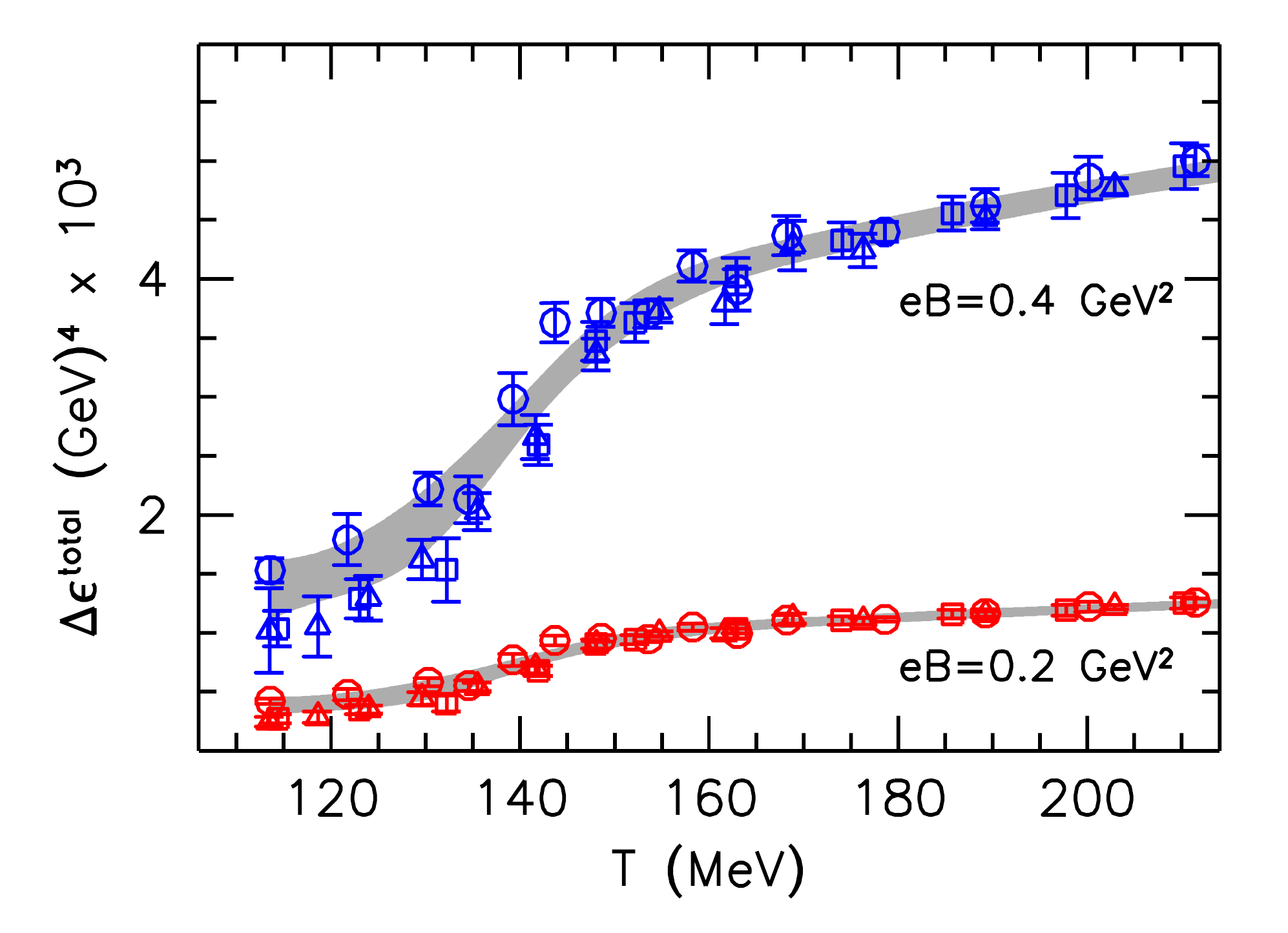}
\vspace*{-.3cm}
\caption{\label{fig:eospI}
Change in the EoS due to the magnetic field. Shown are the longitudinal pressure 
(upper left panel), the $\Phi$-scheme interaction measure (upper right panel), 
the magnetization (lower left panel) and the energy density (lower right panel) 
as functions of the temperature for three lattice spacings and two values of $eB$. 
The shaded areas correspond to our continuum estimates (see the text).
}
\vspace*{-.4cm}
\end{figure}

At $B=0$ the usual normalization of, e.g., the pressure is $p/T^4$. 
In our case this may not be the optimal choice, since $p$ contains terms of 
$\mathcal{O}((eB)^4)$ at zero temperature, which give rise to a $\sim1/T^4$ 
divergence towards $T=0$. We demonstrate this in Fig.~\ref{fig:eos_norm} for the 
case of the longitudinal pressure (left panel) and the interaction measure in the $\Phi$-scheme (right panel). 
It is therefore instructive to plot the observables without this normalization with respect to 
$T^4$. First we show 
the change in the EoS induced by the magnetic field, $\Delta p_{z,r}$ and $\Delta I^{(\Phi)}_r$ for 
two values of $eB$, see Fig.~\ref{fig:eospI}. 
We find that in the low-temperature region our three lattice spacings do not 
suffice to perform a controlled continuum extrapolation\footnote{A well-known source 
of lattice artefact is the taste symmetry breaking of staggered fermions, which 
may lead to large discretization effects at low temperatures. 
Note that not all observables are affected equally by these artefacts.}. 
To allow for a parameterization of our 
results we consider a continuum estimate as the average of our $N_t=8$ and $N_t=10$ results. 
This estimate is indicated by the gray shaded bands in the figures and is used in the following analysis.
Note the positivity of $\M$, which indicates the paramagnetic nature of the 
thermal QCD medium for the whole temperature range.

\begin{figure}[b]
\centering
\includegraphics[width=7.8cm]{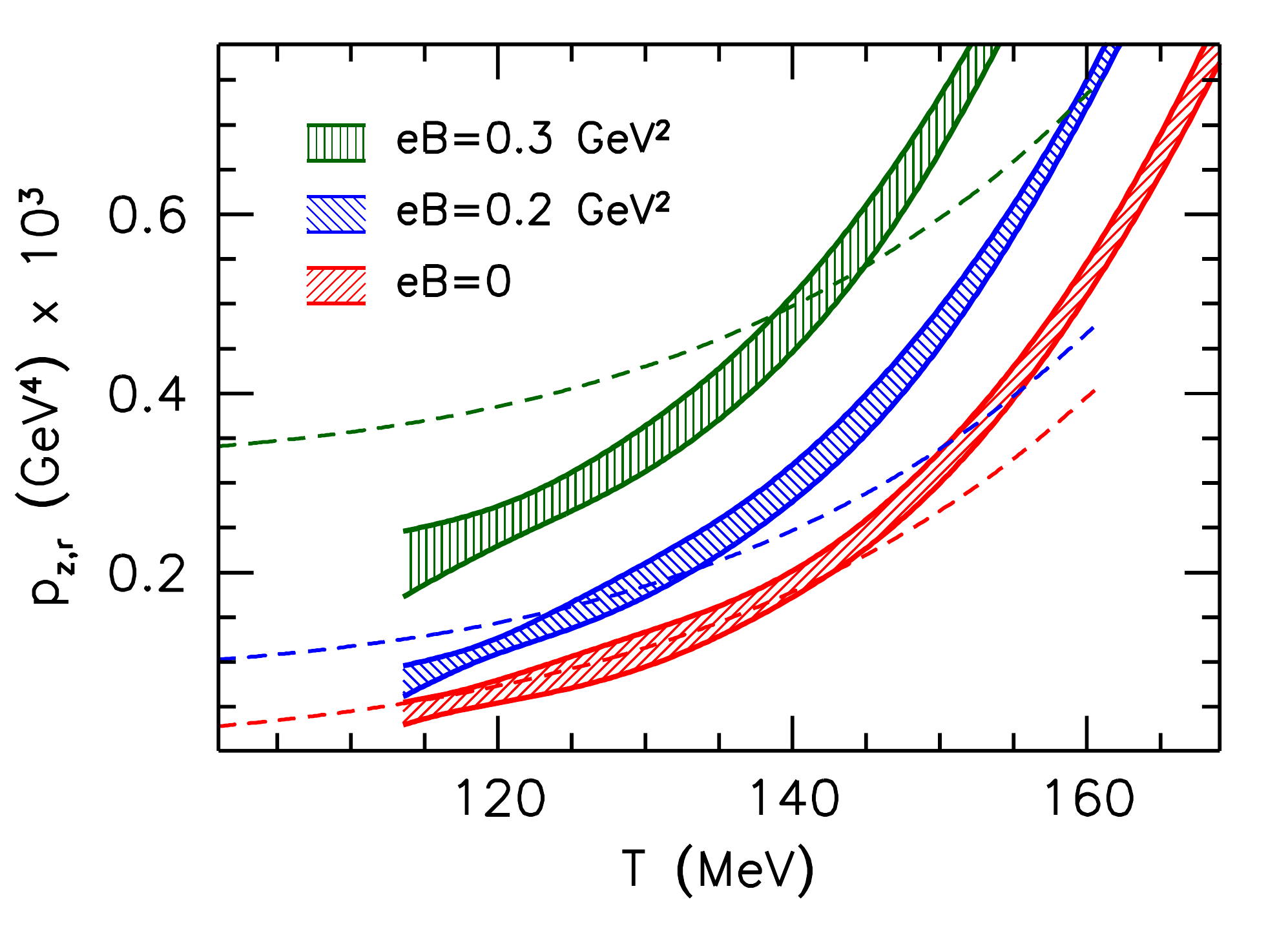} \quad\quad
\includegraphics[width=7.8cm]{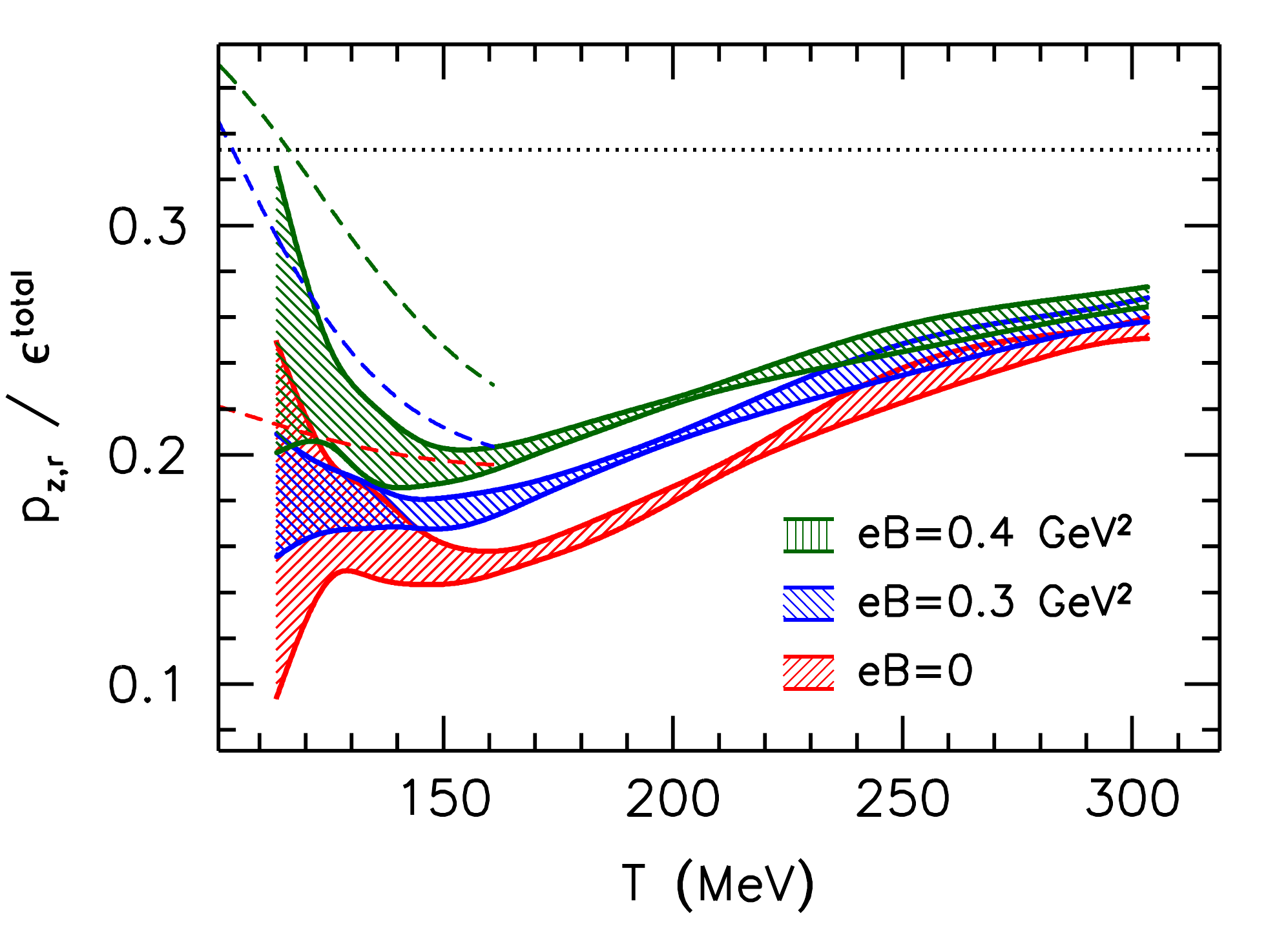}
\vspace*{-.3cm}
\caption{\label{fig:cont_total}
Longitudinal pressure (left panel) and 
the ratio of pressure over energy density (right panel) 
as functions of the temperature, for various values of the magnetic field (note the 
different $T$-ranges). The shaded bands 
indicate the continuum estimates from the lattice results and the dashed lines correspond to 
the HRG model prediction. The dotted line for $p_{z,r}/\epsilon^{\rm total}$ 
signals the Stefan-Boltzmann limit $1/3$.
}
\end{figure}

We proceed by discussing the full EoS at nonzero magnetic fields and concentrate on the low-temperature 
region where the HRG model is expected to be a valid description. 
In Fig.~\ref{fig:cont_total} we show the longitudinal pressure and the ratio of pressure 
over total energy density. 
Comparing $p_{z,r}$ with the HRG model reveals that the model 
overestimates the pressure at nonzero magnetic fields. This mismatch between the model and the lattice 
results was not yet visible in our previous comparison~\cite{Bali:2013txa}, where only the $N_t=6$ 
lattice data were available. 
The ratio $p_{z,r}/\epsilon^{\rm total}$ (see Eq.~(\ref{eq:fundeq}) for our definition of $\epsilon^{\rm total}$) exhibits a shallow dip in the transition region, which 
moves towards the left, signaling the reduction of the transition temperature 
as $B$ grows, in accordance with our earlier findings with other thermodynamic observables~\cite{Bali:2011qj,Bruckmann:2013oba}. 
This dip is, however, not pronounced enough to enable us to reliably determine the position 
of its minimum. 
We will study the dependence of the transition temperature on $B$ in Sec.~\ref{sec:phasediag}.

\begin{figure}[ht!]
\centering
\includegraphics[width=7.8cm]{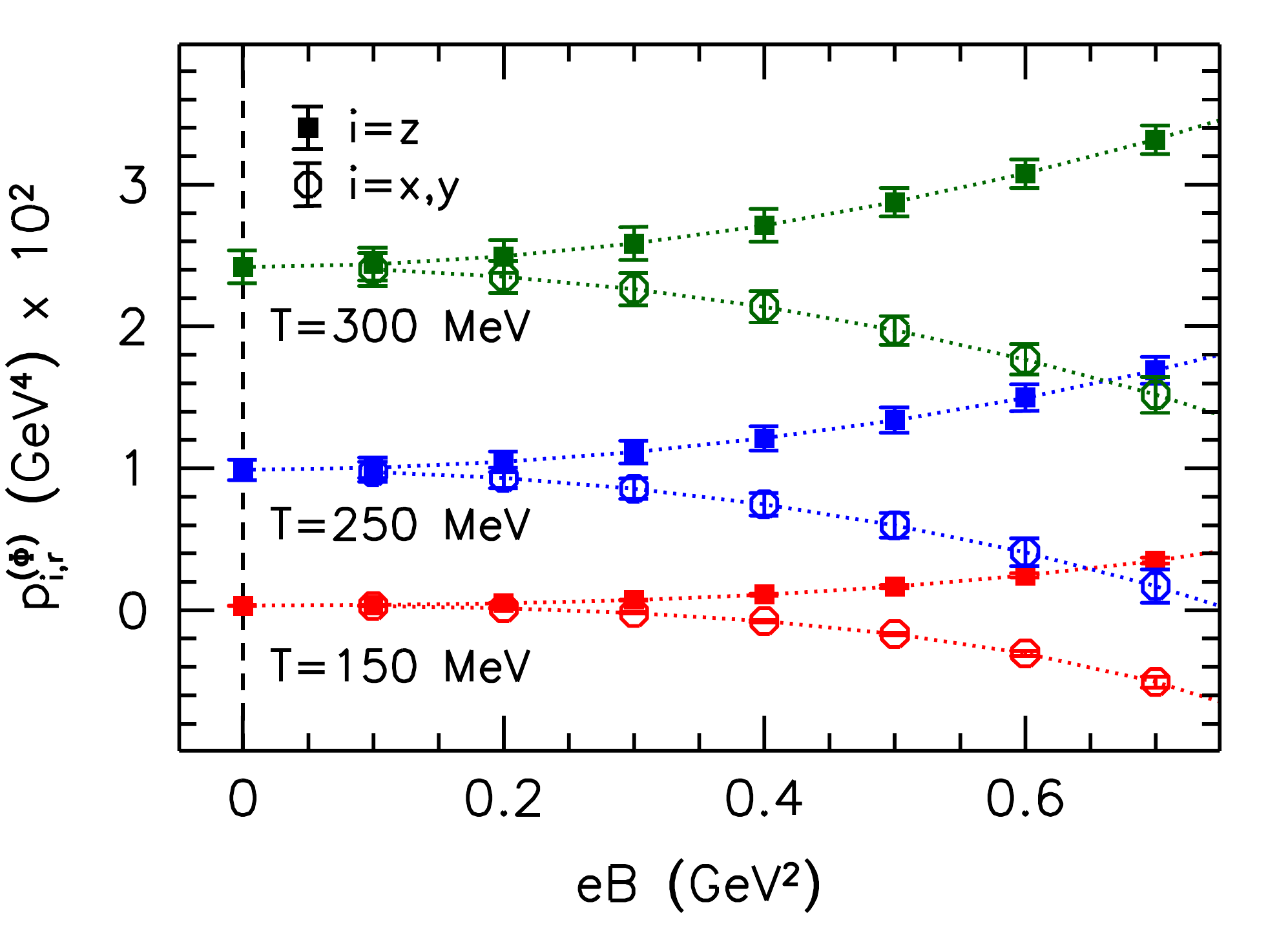} \quad\quad
\includegraphics[width=7.8cm]{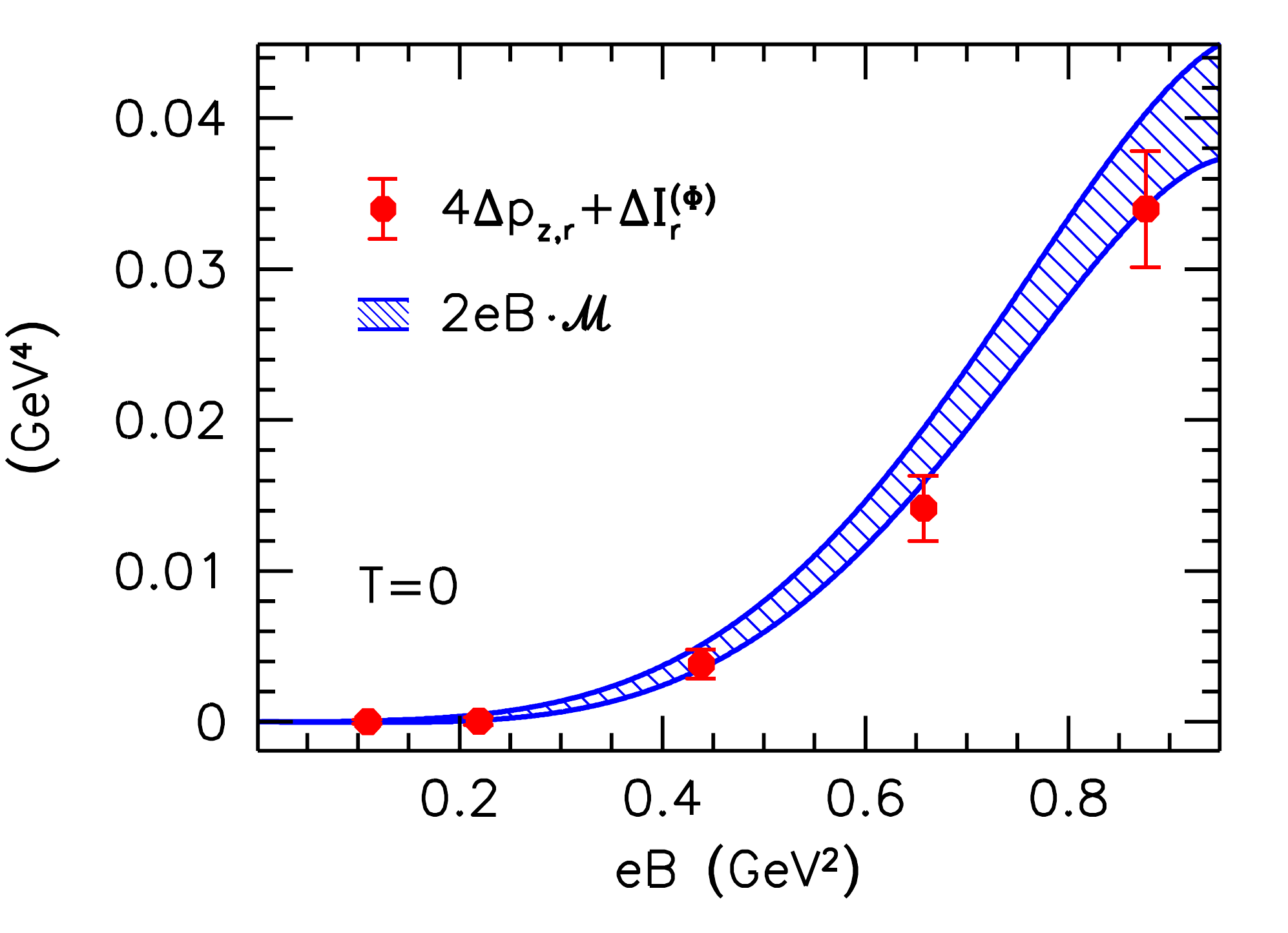}  
\vspace*{-.3cm}
\caption{\label{fig:splitting}
Left panel: splitting of the $\Phi$-scheme pressure components due to the magnetic field for 
a few temperatures. The upper branches correspond to $p^{(\Phi)}_{z,r}$, whereas 
the lower ones to $p^{(\Phi)}_{x,r}=p^{(\Phi)}_{y,r}$.
Right panel: consistency check at $T=0$ (see the text).}
\end{figure}

Up to now we only discussed the longitudinal pressure. Depending on which scheme is used, 
the magnetic field can induce an anisotropy between the pressure components, see Eq.~(\ref{eq:BPHI}). 
The $\Phi$-scheme describes a situation, in which the transverse compression of the 
system proceeds at fixed magnetic flux, thereby inducing an anisotropy that is 
proportional to the magnetization. 
This scheme is adequate for systems where the 
magnetic field lines are frozen in the medium -- i.e.\ where the electric conductivity 
is infinite. 
The parallel and perpendicular components of the $\Phi$-scheme pressure are shown 
in the left panel of Fig.~\ref{fig:splitting}. 
The splitting between the components grows as $T$ increases, 
due to the logarithmic rise of the magnetization, cf.\ Sec.~\ref{sec:perme} below. 
Note that for large magnetic fields the transverse pressure components become 
negative. This is due to the positivity of $\M$, which implies that the free energy 
decreases with growing $B$, i.e.\ the system prefers large magnetic 
fields. For fixed $\Phi$, this preference leads to a collapse of the medium 
in the transverse directions and is signalled by the negative transverse 
pressure. This unphysical instability invalidates the $\Phi$-scheme for large 
magnetic fields, and is avoided if a finite (physical) electric conductivity is 
considered, such that magnetic field lines are not completely frozen 
and magnetic flux is not completely conserved. We emphasize again that the notion of transverse pressure 
depends on its precise definition (i.e.\ the scheme) and should be specified 
for each problem in question.
Our lattice results for $p_z$ and for $\M$ are reliable for the whole magnetic field range under study, 
and can be combined to obtain the transversal pressures in an arbitrary ``general'' scheme 
according to Eq.~(\ref{eq:generalscheme}).

Since the above described analysis to obtain the equation of state involves several 
interpolations, we perform one additional consistency check at $T=0$. 
Here the longitudinal pressure and the energy density coincide (the $t$- and $z$-directions 
are indistinguishable even in the presence of the magnetic field, 
i.e.\ $f_r = -p_{z,r} = \epsilon$
), which gives a relation 
between the trace of the energy-momentum tensor (the interaction measure in the $\Phi$-scheme) 
and the pressures. For the renormalized quantities this reads
\be
T=0:\quad\quad\quad 4 \Delta p_{z,r}  + \Delta I^{(\Phi)}_r = 2 \eBM.
\label{eq:consistency}
\ee
The left hand side of this relation can be obtained without new inter/extrapolations. The 
renormalization involves subtracting $4[c_0+b_1^{\rm free} \log(a/a_0)] \cdot (eB)^2$ for 
the pressure part and $b_1^{\rm free} \cdot(eB)^2$ for the interaction measure part, respectively 
(the necessary parameter values are listed in Table~\ref{tab:chren}). 
The right hand side is obtained by interpolating and differentiating the renormalized longitudinal 
pressure (as was already done to obtain $\M$ of Fig.~\ref{fig:eospI} at $T>0$), cf.\ Eqs.~(\ref{eq:diffrels}) and~(\ref{eq:pzdef}). 
The two sides are compared in the right panel of Fig.~\ref{fig:splitting} for a zero-temperature 
lattice at $\beta=3.55$ (corresponding to $a=1.09 \textmd{ GeV}^{-1}$), showing nice 
agreement. Note that the relation Eq.~(\ref{eq:consistency}) is only valid at vanishing 
temperature and is subject to corrections as $T$ grows. 

\subsection{Magnetic susceptibility and permeability}
\label{sec:perme}

\begin{figure}[ht!]
\centering
\includegraphics[width=7.8cm]{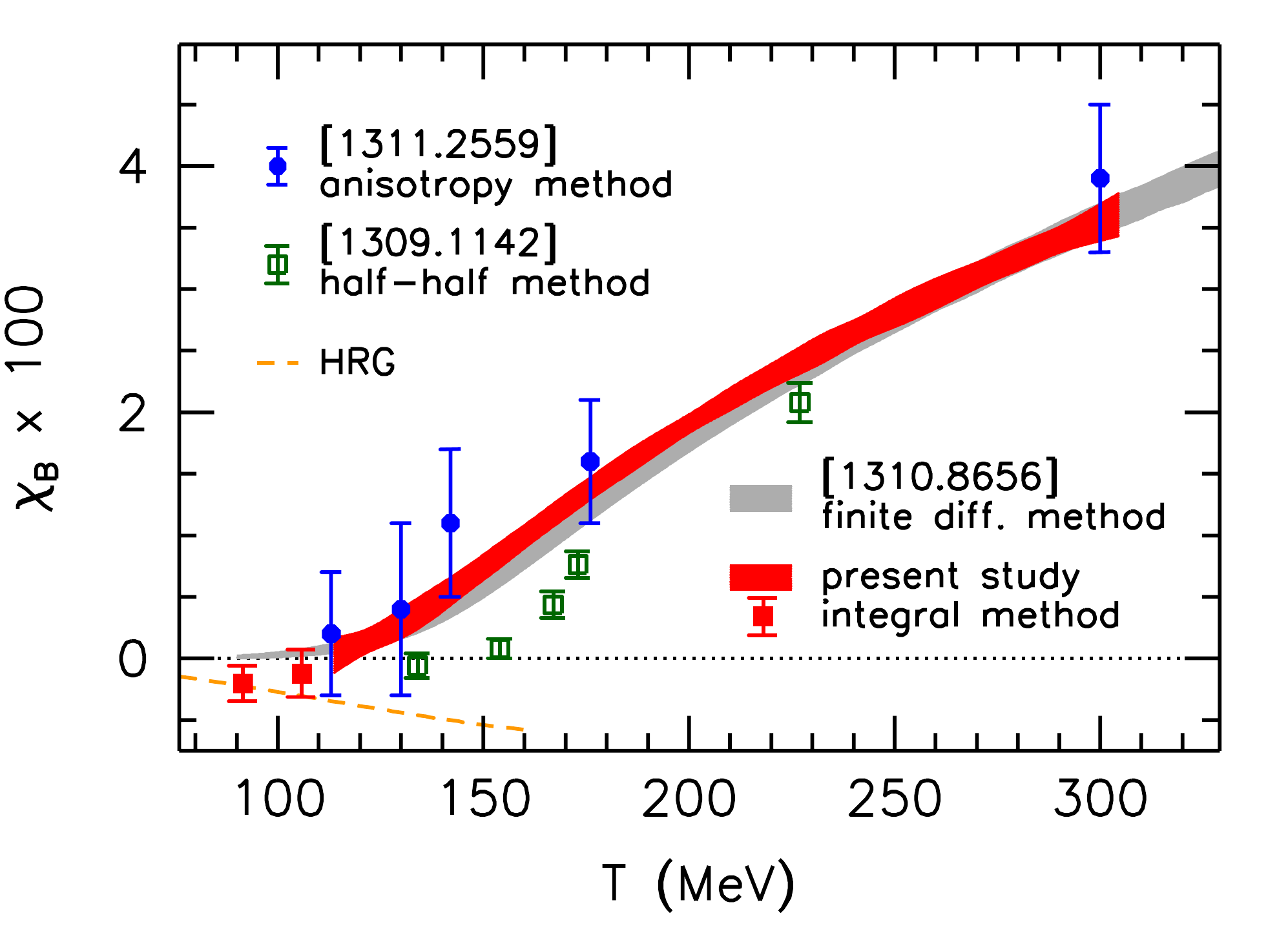} \quad\quad
\includegraphics[width=7.8cm]{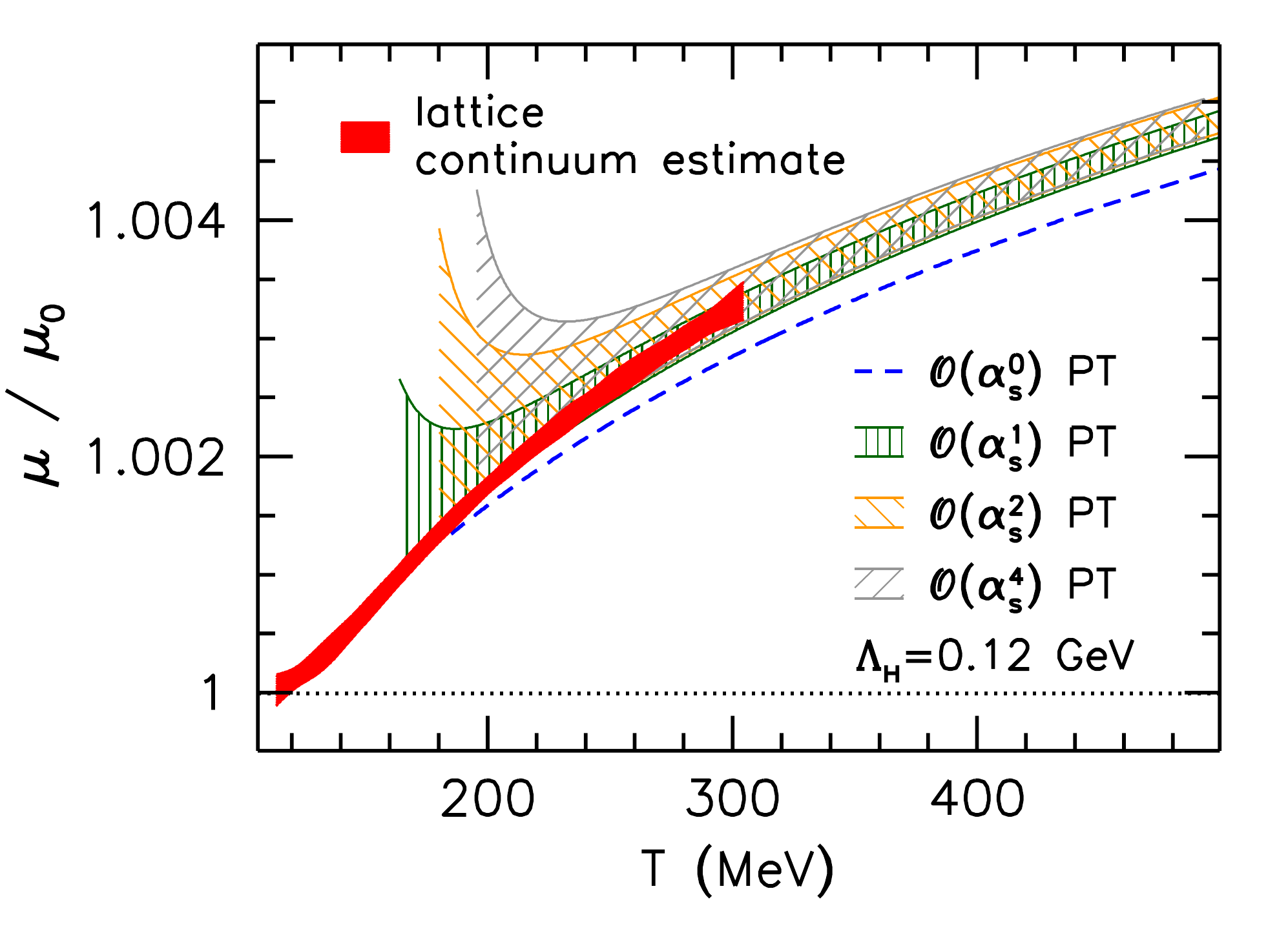}
\vspace*{-.3cm}
\caption{\label{fig:susc}
Left panel: magnetic susceptibility of QCD as a function of the temperature. Results with 
different lattice approaches are collected. Right panel: QCD magnetic permeability 
in units of the vacuum permeability $\mu_0$, and a comparison to perturbation theory,
truncated at various 
orders of the strong coupling. }
\end{figure}

The low-$B$ behavior of the magnetization provides the magnetic susceptibility $\chi_B$, 
as defined in Eq.~(\ref{eq:defsusc}). 
In the left panel of Fig.~\ref{fig:susc} we show our results for $\chi_B$, compared to those 
obtained with various other methods (see the discussion in Sec.~\ref{sec:fluxq}). 
The new results agree within errors with our previous results employing the anisotropy 
method~\cite{Bali:2013owa} and with the results of Ref.~\cite{Bonati:2013vba} using 
the finite difference method. In these studies the same lattice action (stout 
improved staggered quarks with physical quark masses, $m_s/m_{ud}=28.15$) were employed, 
and the data in each case correspond to a similar continuum estimate as we discussed above. 
We stress that the three approaches are completely different, but nevertheless show excellent 
agreement.
We also include the susceptibility obtained in Ref.~\cite{Levkova:2013qda}, where the HISQ 
action with nearly physical quark masses ($m_s/m_{ud}=20$) was used and the half-half method 
was employed on $N_t=8$ lattices. The susceptibility is observed to be somewhat smaller than in the other approaches, 
which may be related to the larger value of the quark mass. 
In summary, all lattice results indicate that the susceptibility is positive and increases 
as $T$ grows, thus signalling the paramagnetic nature of the thermal QCD medium for 
temperatures around and above the transition region.

The susceptibility can also be calculated using the HRG model. Here the EoS is represented 
by a sum over contributions of non-interacting hadrons and resonances. At low temperatures 
and magnetic fields, the dominant term in the sum is given by the lightest 
hadrons, i.e.\ pions. The pionic contribution to $\chi_B$ is negative (see the discussion 
in App.~\ref{app:chiHRG}), suggesting the presence of a (weakly) diamagnetic region at 
low $T$.
Contrary to pions, higher-spin hadrons give positive contributions to the susceptibility, 
such that $\chi_B$ is expected to bend back towards positive values as $T$ 
grows. 
The HRG prediction is plotted in the left panel of Fig.~\ref{fig:susc}, indeed exhibiting 
a negative region at low temperatures.
Note that at $\mathcal{O}((eB)^4)$, the HRG magnetization receives contributions from 
the $T=0$ vacuum term as well, which are positive both for pions and for spin-$1/2$ and 
spin-$1$ hadrons. Eventually, this results in a positive magnetization for magnetic fields 
exceeding $eB\approx 0.05\textmd{ GeV}^2$ for all temperatures~\cite{Endrodi:2013cs}, in 
agreement with our lattice results in the lower left panel of Fig.~\ref{fig:eospI}. 

In order to extend the temperature range of the lattice results, we include 
two more sets of $28^3\times 10$ lattice ensembles at $T=90\textmd{ MeV}$ ($\beta=3.55$) and at 
$T=105 \textmd{ MeV}$ ($\beta=3.6$). We make use of Eq.~(\ref{eq:pres1}) to obtain the magnetic 
field-dependence of the pressure, which directly gives the susceptibility. The result is indeed 
found to be consistent with the HRG prediction, see the red squares 
in the left panel of Fig.~\ref{fig:susc}. At $T=90 \textmd{ MeV}$ we also measure $\chi_B$ using the 
half-half method, and use the $24^3\times 32$ ensemble for renormalization. 
Consistent with the above lattice determination, we find $\chi_B(T=90\textmd{ MeV})=-0.002(2)$. 
Altogether, the lattice results are compatible with the existence of a weakly diamagnetic region at low 
temperatures.
Our conclusion from this comparison is that the HRG model correctly captures the 
diamagnetic effect of pions, but overestimates their role for $T\gtrsim 120 \textmd{ MeV}$. 
Note that a negative magnetization has also been obtained within the Parton-Hadron-String 
Dynamics approach~\cite{Steinert:2013fza}. This model, however, predicts that 
the diamagnetic response persists to higher magnetic fields ($eB\approx 0.2-0.7 \textmd{ GeV}^2$) 
and is even enhanced as the temperature grows. This is in conflict with both the lattice results 
and the HRG model description presented above. 
The magnetic susceptibility has also been calculated in a model with free quarks coupled 
to the Polyakov loop, giving results consistent with the lattice data at high temperatures~\cite{Orlovsky:2014kva}.

The susceptibility can also be translated to 
the magnetic permeability $\mu$ of the thermal QCD medium. To write down the relation between $\chi_B$ 
and $\mu$, we need to distinguish between the magnetic induction $B^{\rm ind}$ 
and the external field $B^{\rm ext}$ that would be present in the absence of the medium. 
The two fields are connected by the magnetization~\cite{landau1995electrodynamics}, 
\be
B^{\rm ind}=B^{\rm ext}+\M\cdot e.\label{eq:HandB}
\ee
In the present study the magnetic field corresponds to $B^{\rm ind}$, since it is the 
field that traverses the lattice and that quarks couple to, such that $\M(B^{\rm ind}\approx0)=\chi_B \cdot eB^{\rm ind}$. 
Using this, the external field $B^{\rm ext}$ can be found from Eq.~(\ref{eq:HandB}). 
Reinserting the factors of $e=\sqrt{4\pi\alpha_{\rm em}}$ in the definitions of $\chi_B$ and 
of $\M$, the magnetic permeability reads
\be
\mu\equiv\frac{B^{\rm ind}}{B^{\rm ext}} = \frac{1}{1- 4\pi\alpha_{\rm em} \cdot \chi_B}.
\ee
In the SI system this is the magnetic permeability in units of the vacuum permeability $\mu_0$, 
cf.\ Ref.~\cite{Bonati:2013lca}. 

In the right panel of Fig.~\ref{fig:susc} we plot $\mu/\mu_0$, and 
compare it to perturbation theory. We discuss some details of this perturbative 
expansion in the following. 
It turns out that -- even for the lowest-order perturbative expansion of the susceptibility -- 
a non-perturbative parameter is necessary to carry out the comparison with the lattice results 
in a consistent manner. This parameter is the scale $\Lambda_{\rm H}$, which plays the role of the renormalization scale, $\mu=\Lambda_{\rm H}$, see the discussion of Sec.~\ref{sec:chrenfullqcd}. 
The necessity of using $\Lambda_{\rm H}$ in this comparison 
is related to the entanglement of ultraviolet and 
infrared divergences in the presence of the magnetic field: the behavior of the bare free energy 
at $a\to0$ is identical\footnote{
This may be understood as follows. 
As the temperature increases, $T$ gradually adopts the role of 
the largest scale in the system and replaces the scale $\Lambda_{\rm H}$ in Eq.~(\ref{eq:T0fB}). 
The renormalization prescription, however, remains unchanged and still amounts to 
the subtraction at a renormalization scale $\mu=\Lambda_{\rm H}$, as contained in 
Eq.~(\ref{eq:T0fB}). Altogether one obtains, to quadratic order, a logarithmic term 
$\log(T/\Lambda_{\rm H})$, from which Eq.~(\ref{eq:susc_pert}) follows directly.} to that of the renormalized free energy at 
$T\to\infty$.
Thus, the high-temperature susceptibility is again governed by the QED $\beta$-function~\cite{Elmfors:1993wj,Cangemi:1996tp,Bali:2013owa}, 
\be
\chi_B(T) = 2\cdot \QEDb \cdot \log( T/ \Lambda_{\rm H}).
\label{eq:susc_pert}
\ee
Taking $\QEDb=\QEDb^{\rm free}$ and inserting the value  
$\Lambda_{\rm H}=0.12 \textmd{ GeV}$ that we determined 
in Sec.~\ref{sec:quadratic} for physical quark masses, 
we obtain the blue dashed curve in the right panel of Fig.~\ref{fig:susc}. 
To improve this perturbative 
expression, we also take into account QCD corrections to $\QEDb^{\rm free}$ using 
Eq.~(\ref{eq:betacorr}), but this time at the thermal scale $\mu_{\rm th}\sim T$ instead 
of at the lattice regulator $1/a$. To calculate $g^2(\mu_{\rm th})$ we use the four-loop 
running coupling and $\Lambda_{\rm QCD}^{\overline{\rm MS}}=0.34 \textmd{ GeV}$ 
for three-flavor QCD~\cite{Beringer:1900zz}. Considering QCD corrections 
up to various orders in $\alpha_s=g^2/(4\pi)$ we obtain the dashed bands in 
the figure. The width of the bands correspond to the uncertainty of the 
thermal scale, which we allow to vary between $\pi T$ and $4\pi T$. 
This range is physically motivated based on convergence arguments~\cite{Braaten:1995ju}. 
The perturbative expansion\footnote{
We mention that to obtain the perturbative improvement of Eq.~(\ref{eq:susc_pert}) 
we only considered the QCD 
corrections in the coefficient of the leading logarithm and ignored possible 
corrections that arise in the sub-leading constant terms (i.e.\ those that could 
modify the dependence on the renormalization scale $\Lambda_{\rm H}$). 
These could be accounted for by considering the scale $\mu_{\rm th}$ (proportional to $T$)
instead of $T$ in \protect Eq.~(\ref{eq:susc_pert}). 
Note moreover that both the higher-order corrections to the $\beta$-function and 
$\Lambda_{\rm H}$ depend on the renormalization scheme, but this dependence must cancel in 
the susceptibility, being a physical observable.
} seems to show a fast convergence even for reasonably 
low temperatures, and agrees nicely with the lattice data in the temperature 
region $200\textmd{ MeV}<T<300\textmd{ MeV}$.

It is important to stress that $\Lambda_{\rm H}$ appeared in the perturbative description due to 
its role as the renormalization scale. Since the susceptibility contains an ultraviolet 
divergence, its renormalization inevitably introduces an ambiguity, expressed as 
a dependence on the renormalization scale. 
To derive Eq.~(\ref{eq:susc_pert}) we relied on two observations: 
that the zero-temperature free energy to $\mathcal{O}((eB)^2)$ is determined 
exclusively by the QED $\beta$-function and that 
at high temperatures $T$ replaces the regulator $1/a$ in this expression. 
Eq.~(\ref{eq:susc_pert}) can be explicitly checked in the free case~\cite{Elmfors:1993wj,Cangemi:1996tp,Bali:2013owa}. 
Note that our arguments about charge renormalization only relate to the $\mathcal{O}((eB)^2)$ 
contributions, whereas the full, $B$-dependent free energy contains much more information 
and is also considerably more complicated. 
Its calculation to $\mathcal{O}(\alpha_s)$ was performed recently using 
the lowest-Landau-level approximation, valid for large magnetic fields~\cite{Blaizot:2012sd}.

\subsection{Entropy density and the Adler function}

As emphasized in Sec.~\ref{sec:renormEoS}, most observables contain magnetic field-induced 
contributions at $T=0$, making the usual normalization (e.g., with respect to $T^4$ for the 
pressure) disadvantageous. 
In this respect, the entropy density is special: being the derivative of $p_{z,r}$ with respect 
to the temperature, it vanishes identically at $T=0$. 
This may be understood from the fact that the vacuum contribution 
is a pure quantum effect (it emerges from the interaction of virtual quarks with the external 
field) and thus it cannot produce entropy. 
Note that at $T>0$, the magnetic field changes the thermal distribution and is expected to modify $s$.

\begin{figure}[ht!]
\centering
\includegraphics[width=7.8cm]{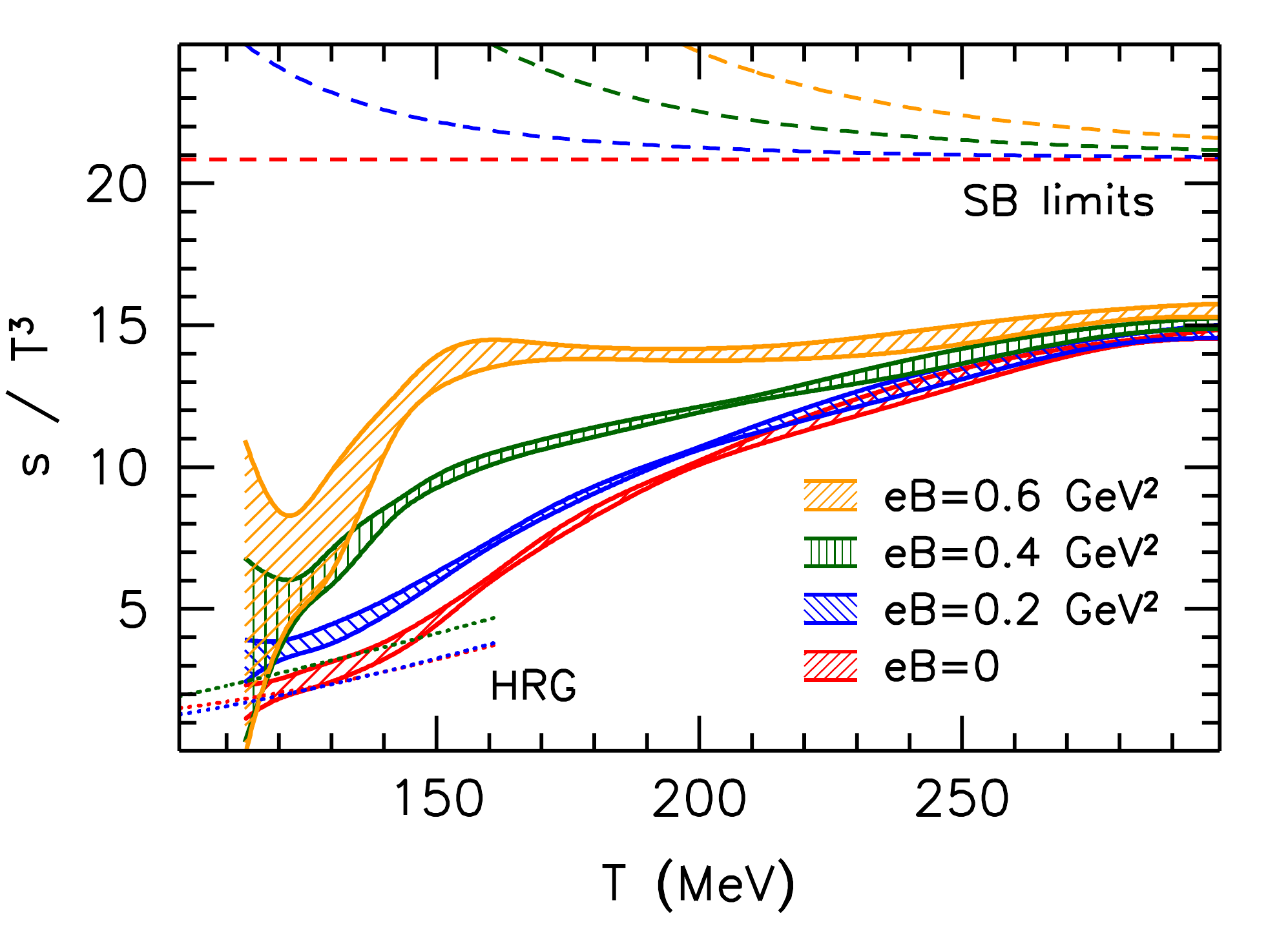} \quad\quad
\includegraphics[width=7.8cm]{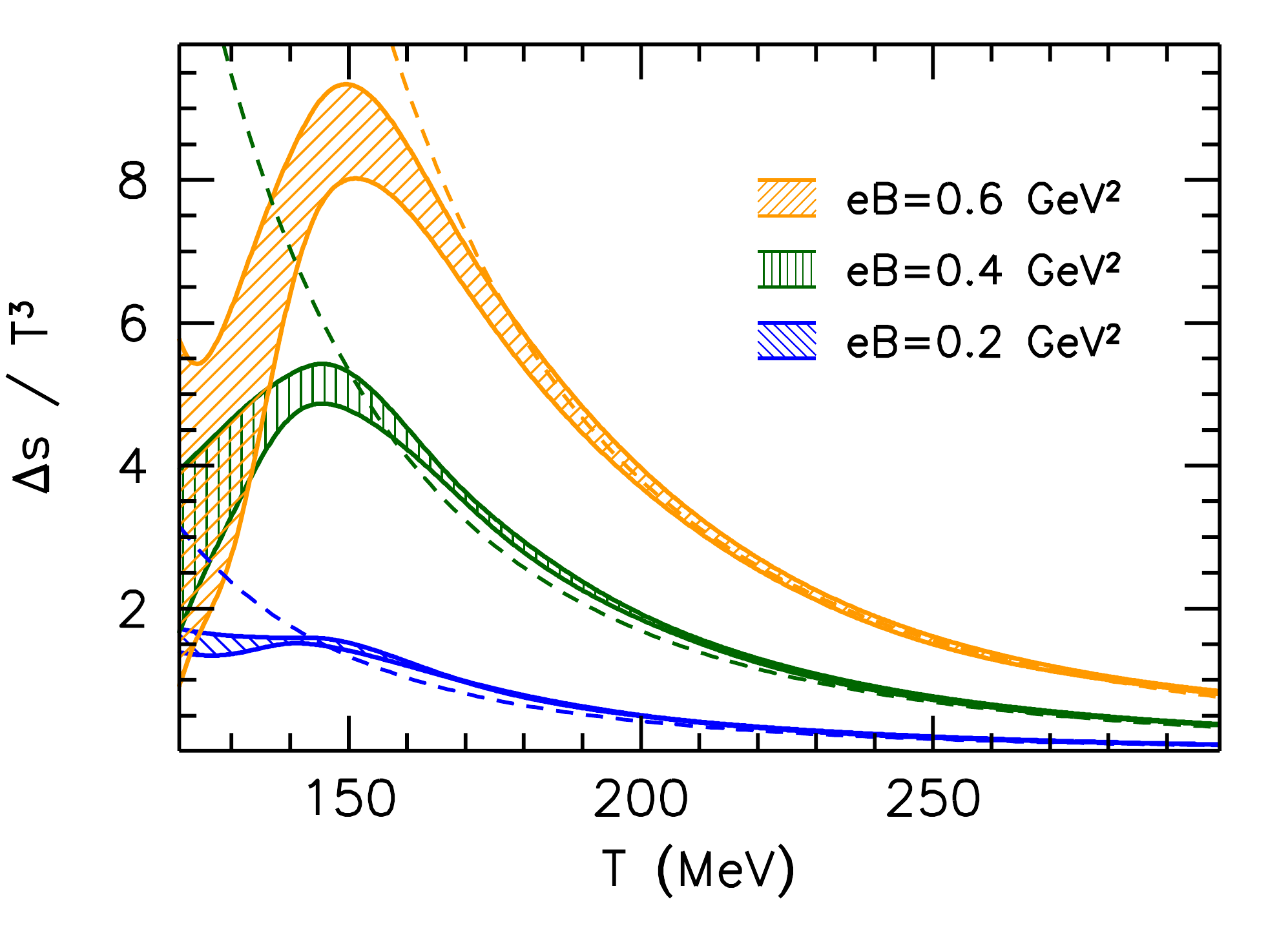}
\vspace*{-.3cm}
\caption{\label{fig:norm_ent}
Left panel: entropy density normalized by $T^3$ for various magnetic fields. The lattice results are 
compared to the HRG prediction (dotted curves at low $T$) and the corresponding 
Stefan-Boltzmann limits (see Eq.~(\ref{eq:sSB})) are also indicated (dashed curves). Right panel: 
normalized entropy density minus its $B=0$ value, compared to the Stefan-Boltzmann limits. 
}
\end{figure}

The vanishing of $s(T=0)$ allows for a study of the usual normalization $s/T^3$, plotted in 
the left panel of Fig.~\ref{fig:norm_ent}. Note that the errors become large at low temperatures as $B$ increases, due to 
the cancellation of the $\mathcal{O}(T^{-3})$ magnetic field-induced contributions in $s$. 
For the lowest three magnetic fields at low temperatures, we also show the HRG prediction in the plot.
Moreover, we indicate the Stefan-Boltzmann limits for each value of $eB$. These can be calculated 
from the dependence of the free pressure on $T$ and $B$,
\be
p_{z,r}^{\rm free} = -f_r = \frac{19\pi^2}{36} T^4 + \QEDb^{\rm free} (eB)^2 \log(T/\Lambda_{\rm H}) + \ldots \quad \to \quad
\frac{s^{\rm free}}{T^3} = \frac{19\pi^2}{8} + \QEDb^{\rm free} \frac{(eB)^2}{T^4} + \ldots,
\label{eq:sSB}
\ee
where we considered three massless quark flavors and used the high-temperature limit of the 
magnetic susceptibility from Eq.~(\ref{eq:susc_pert}). Note that the renormalization scale 
$\Lambda_{\rm H}$ cancels in the entropy density. In the right panel of Fig.~\ref{fig:norm_ent} 
we show the magnetic field-induced part $\Delta s/T^3$, compared to the expected perturbative 
behavior. 
Note that while $s(B=0)$ is by $20-30\%$ below its Stefan-Boltzmann limit at our highest temperature, 
$\Delta s$ almost perfectly follows the free-case prediction, already for $T\gtrsim 170 \textmd{ MeV}$. 
A similar behavior is observed for the pressure and, thus, for all other observables as well. 
In other words, the $B$-dependence of the EoS in the deconfined phase is 
predominantly dictated by the $\mathcal{O}(B^2)$ free-case behavior -- in sharp contrast 
to the $B=0$ EoS, which shows strong deviations from the perturbative predictions within this 
range of temperatures. Similar conclusions were drawn using a perturbative treatment of QCD in magnetic fields, where the $\mathcal{O}(g^2)$ term was shown to be suppressed with respect to the free-case contribution~\cite{Blaizot:2012sd}.

The correspondence between the entropy density and perturbative QED physics 
can be pushed even further: we find that the second derivative of the entropy density with respect to the 
magnetic field at $B=0$ is related 
to the Adler function (for its definition, see, e.g., Ref.~\cite{Baikov:2010je}) 
at high temperatures. To understand the origin of this relation, it is 
advantageous to consider a
(perturbative) diagrammatic representation of the 
free energy $f(B)$:
it consists of all closed loop diagrams 
containing virtual quarks and
\begin{wrapfigure}{r}{8.2cm}
\centering
\vspace*{-.3cm}
\includegraphics[width=7.8cm]{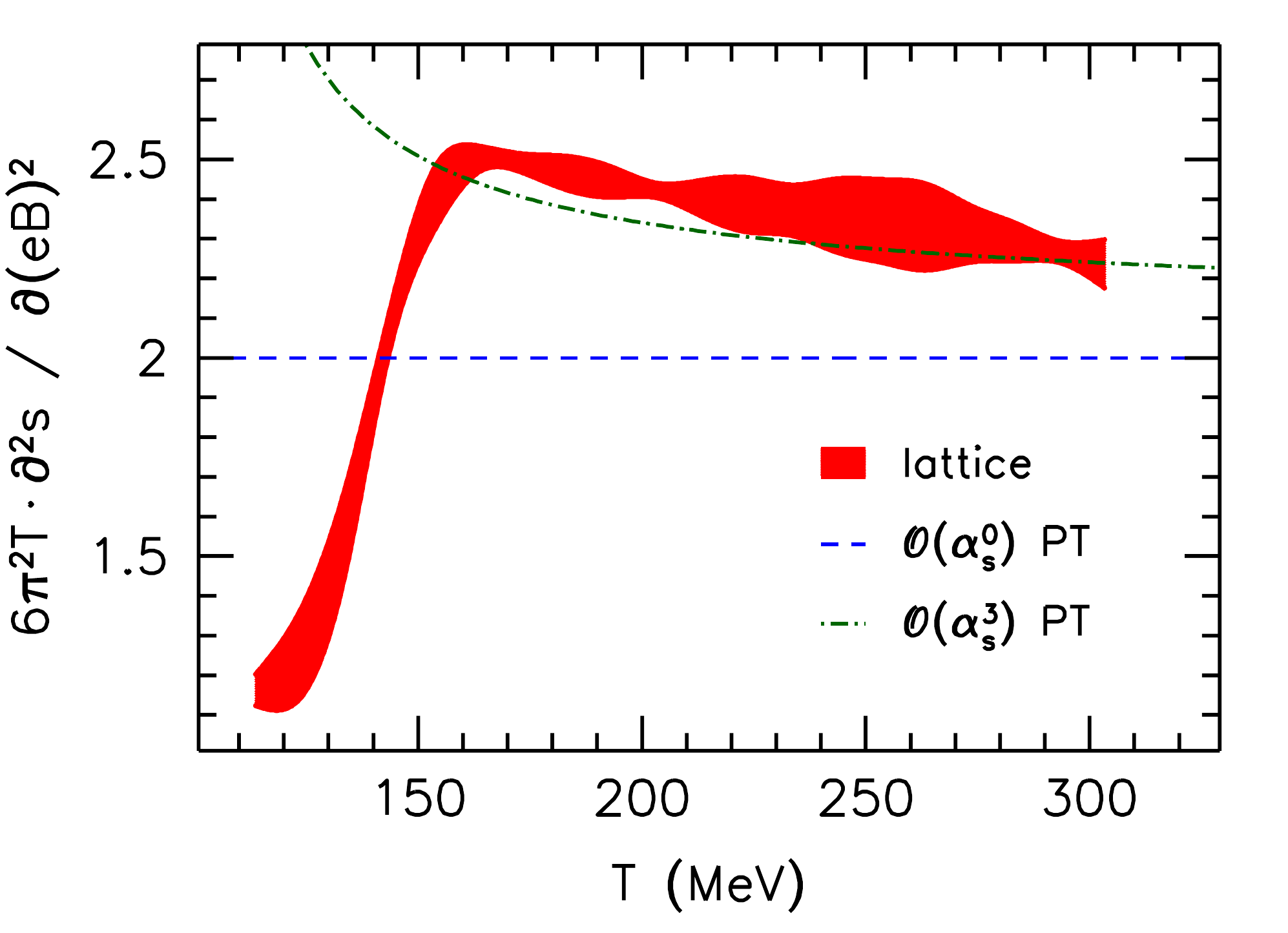}
\vspace*{-.3cm}
\caption{\label{fig:adler}
Second derivative of the entropy density with respect to $eB$ at high temperatures, 
compared to perturbation theory. 
}
\end{wrapfigure}
gluons. On the one hand, taking the second derivative 
with respect to $eB$ gives the susceptibility $\chi_B$. On the other hand, this 
derivative effectively pulls out two photon legs (these photons correspond to the background magnetic field), giving the 
photon vacuum polarization diagram $\Pi$. This diagram is usually considered with an 
inflowing momentum $Q$, and the 
Adler function is defined from it via $D(Q) = 12\pi^2\cdot\partial \Pi/\partial \log Q^2$. The equivalent 
of the momentum $Q$ in our setup is expected to be the largest scale in the system: at high 
temperatures, for example, $Q\sim T$ (remember that to define $\chi_B$, the magnetic 
field has already been set to zero). Thus, in this region we expect that the Adler function is 
approached as
\be
D(\mu_{\rm th}) \quad \longleftrightarrow \quad 12\pi^2 \cdot\frac{\partial \chi_B}{\partial \log T^2} = 
6\pi^2 \, T\cdot\frac{\partial^2 s}{\partial (eB)^2},
\label{eq:adler}
\ee
where we interchanged the derivatives with respect to $T$ and to $eB$. 
Eq.~(\ref{eq:adler}) reveals a relation between the magnetic field-dependence 
of the entropy and the Adler function at a thermal scale $\mu_{\rm th}\sim T$. 
In Fig.~\ref{fig:adler} we show our continuum estimate for the right-hand-side of Eq.~(\ref{eq:adler})
and a comparison to the perturbative expansion~\cite{Baikov:2010je} of $D(\mu_{\rm th})$
(where we used $\mu_{\rm th}=2\pi T$).
Note that the above correspondence is only expected to be valid for $T \gg \Lambda_{\rm H}$, 
where the relevant scale is uniquely defined by the temperature. 
Note also that according to this argument, Eq.~(\ref{eq:adler}) fixes the asymptotic dependence of $\partial^2 s/\partial(eB)^2$ 
on any external parameter (e.g., on chemical potentials), as long as this parameter 
represents the largest scale in the system. The correspondence between the susceptibility and the 
Adler function clearly deserves a more detailed investigation, which we plan to conduct 
in the near future.

\subsection{Phase diagram}
\label{sec:phasediag}

\begin{figure}[h]
\centering
\includegraphics[width=7.8cm]{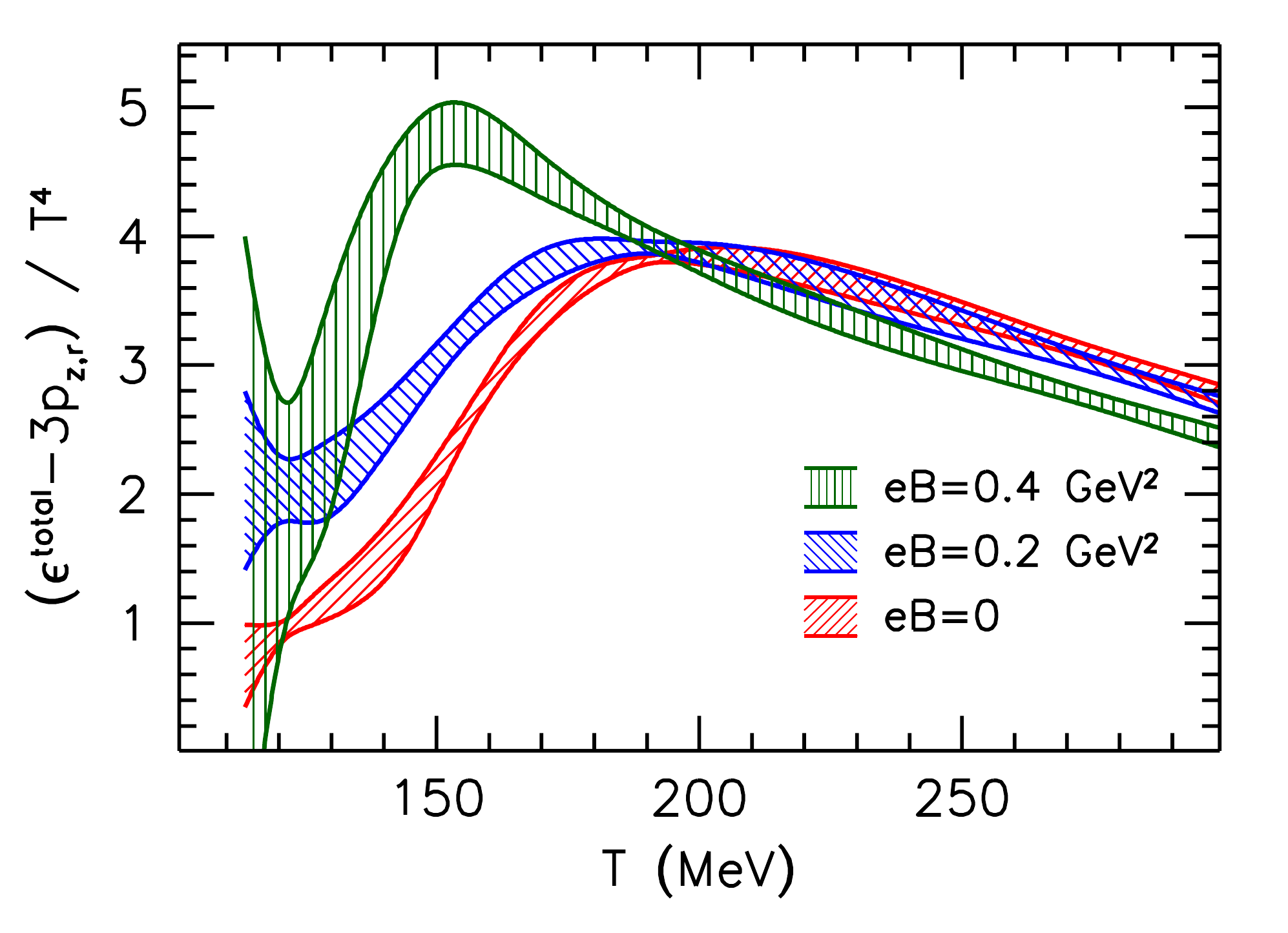} \quad\quad
\includegraphics[width=7.8cm]{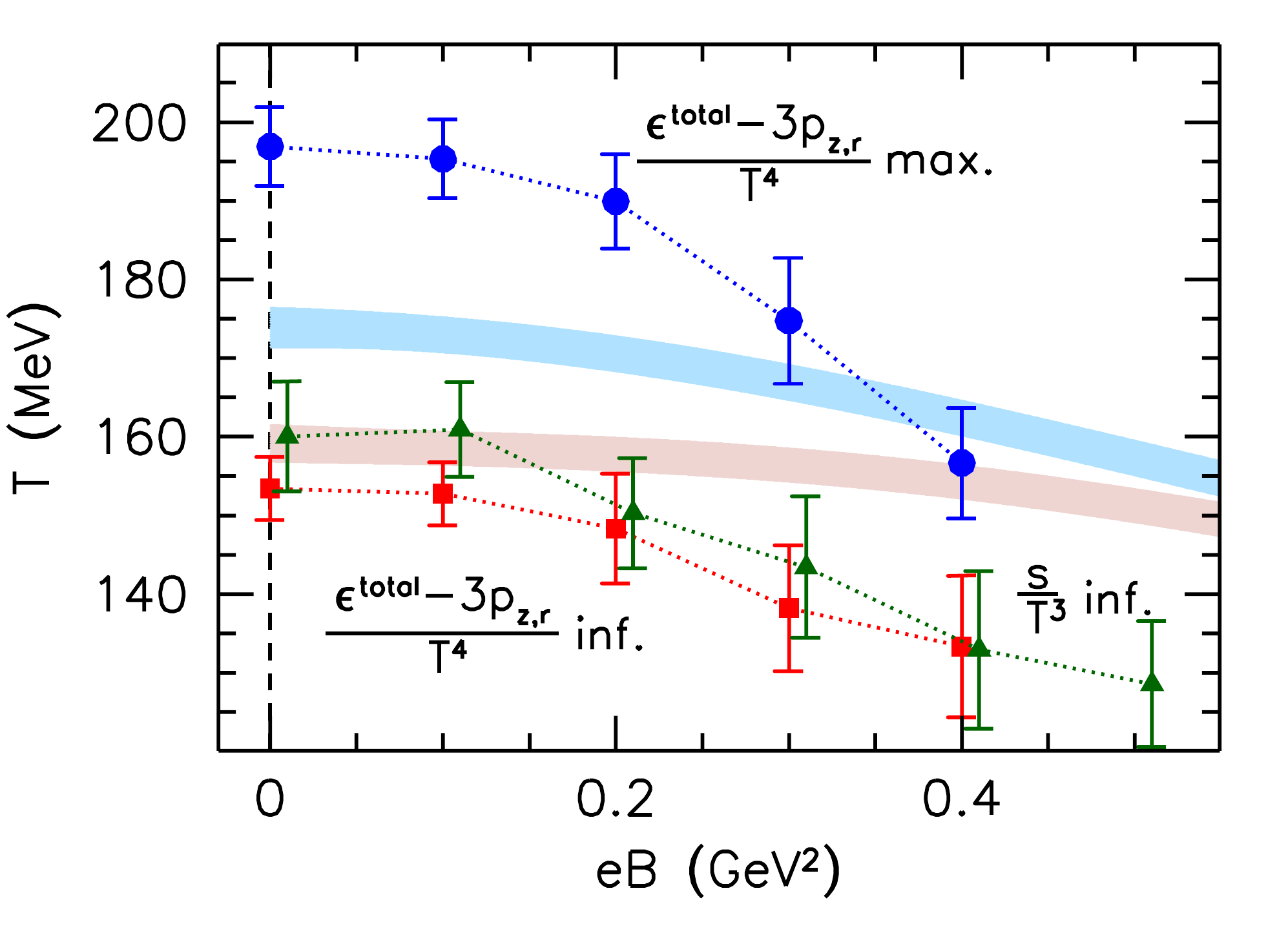}
\vspace*{-.3cm}
\caption{\label{fig:pdI}
Left panel: the combination $(\epsilon^{\rm total}-3p_{z,r})/T^4$ as a function 
of $T$ and $B$. Right panel: magnetic field-dependence of the 
characteristic points (inflection point and maximum) of EoS observables (points), 
compared to our earlier results\protect~\cite{Bali:2011qj} 
for the inflection point of the strange quark number susceptibility (light blue 
shaded band) and that of the light quark condensates (light red shaded band). 
}
\end{figure}

Let us now use the dependence of the EoS on $T$ and on $B$ to discuss the QCD phase diagram 
in the $B-T$ plane. To this end we need to define $T_c(B)$ through characteristic 
points of some observables. We have seen that most observables are nonzero at $T=0$, 
making the usual normalization by $T^4$ disadvantageous. 
The shift in $p_{z,r}$ at zero temperature is of $\mathcal{O}((eB)^4)$ and positive, as we 
have seen in Sec.~\ref{sec:renormEoS}. 
It is simple to show using this observation and the thermodynamic relations 
of Sec.~\ref{sec:thermo} that $\eBM$, $I^{(\Phi)}_r$ and $\epsilon^{\rm total}$ are also 
of $\mathcal{O}((eB)^4)$ and positive at $T=0$. Conversely, $\epsilon$ and $I^{(B)}_r$ 
are of $\mathcal{O}((eB)^4)$ and negative at zero temperature. 
A special combination is $\epsilon^{\rm total}-3p_{z,r} = I^{(\Phi)}_r -\eBM$ for which 
quartic terms also cancel and which is of $\mathcal{O}((eB)^6)$. 
Let us consider this combination in somewhat more detail. It is plotted 
in the left panel of Fig.~\ref{fig:pdI}. This combination coincides with the 
renormalized interaction measure 
at $B=0$. Accordingly, it exhibits a pronounced peak, which 
is observed to move towards lower temperatures as $B$ increases. Eventually, the $\mathcal{O}((eB)^6)$
terms start to contribute to this combination, making it diverge at low temperatures and 
washing out the peak-like structure. Still, up to $eB=0.4 \textmd { GeV}^2$ we can use the peak of this observable to characterize the transition region. 

For a first-order phase transition, the inflection point of this observable would turn 
into a discontinuity, thus this point marks the (pseudo-critical) transition temperature 
$T_c(B)$ for the crossover. Similarly, the inflection point of $s/T^3$ (see the left panel 
of Fig.~\ref{fig:norm_ent}) also represents 
a candidate for defining the transition temperature. 
Together with the maximum of $(\epsilon^{\rm total}-3p_{z,r})/T^4$ (which, however, does not correspond to a 
pseudo-critical temperature, but is merely a characteristic point), we show the $B$-dependence of these 
definitions in the right panel of Fig.~\ref{fig:pdI}, and compare them to our earlier determinations 
of the phase diagram using the strange quark number susceptibility and the light quark 
condensates.
The results consistently show a reduction of the transition temperature 
as $B$ grows. 
Note that the difference between different determinations of $T_c$ reflects the crossover nature of the 
transition. The variance between the four definitions is found to remain constant within the errors.

\section{Summary}
\label{sec:summary}

Using a novel `generalized integral method', we determined the QCD equation of state 
for a wide range of temperatures 
and magnetic fields up to $eB=0.7 \textmd{ GeV}^2$. 
Results were presented for a variety of thermodynamic observables, 
indicating that the EoS is significantly affected by the magnetic field, even at moderate 
values of $B$. 
The tabulated data is available in the ancillary files \texttt{table\_EoS\_B.dat} 
and \texttt{table\_EoS\_B\_derivatives.dat}, 
submitted to the arXiv along with this paper. 

The thermodynamic structure of QCD is altered by the magnetic field in several aspects:
\begin{itemize}
\item {\bf Vacuum term.} 
The magnetic field induces a vacuum contribution to most observables, such that e.g.\ the 
pressure does not vanish at $T=0$. This vacuum term makes the usual normalization of 
the affected observables (e.g.\ $p/T^4$) ill-defined at low temperatures.

\item {\bf Pressure anisotropy.}
The magnetic field creates an anisotropy between the pressure components if 
the transverse pressure is defined at constant magnetic flux ($\Phi$-scheme). This 
anisotropy is sizeable and becomes comparable to the longitudinal pressure as $T$ 
or $B$ increase. 

\item {\bf Para-/diamagnetism.}
The leading-order response of the system to $B$ is characterized by the 
magnetic susceptibility $\chi_B$. On the one hand, the thermal QCD medium 
is a strong paramagnet ($\chi_B>0$) around and above the transition region. On the other hand,
there appears to be a weakly diamagnetic region ($\chi_B<0$) at low temperatures $T\lesssim 100 \textmd{ MeV}$, 
where pions dominate.

\item {\bf Validity of HRG and of PT.}
For the susceptibility, the Hadron Resonance Gas model breaks down already at 
$T\approx 120 \textmd{ MeV}$.
However, perturbation theory successfully describes the lattice data at suprisingly low 
temperatures $T\approx 200-300 \textmd{ MeV}$. 

\end{itemize}

The presence of the background magnetic field necessitates the renormalization of the 
electric charge. This has several implications:

\begin{itemize}
\item {\bf Renormalization.}
The pressure undergoes additive renormalization at $T=0$. The divergent term that needs to be subtracted 
is logarithmic in the lattice spacing and its coefficient equals the lowest-order 
QED $\beta$-function (with QCD corrections at the scale $1/a$).

\item {\bf Magnetic catalysis.}
At zero temperature, the phenomenon of magnetic catalysis
(the enhancement of the quark condensate by the magnetic field) to quadratic order in $eB$ 
is related to the positivity of the QED $\beta$-function. 

\item {\bf Susceptibility.}
For high temperatures, the magnetic susceptibility increases logarithmically with $T$, at a rate 
given by the QED $\beta$-function (with QCD corrections at the scale $T$). 

\item {\bf Adler function.}
The second derivative of the entropy density with respect to the magnetic field 
is related to the perturbative Adler function at high temperatures. 
\end{itemize}

In addition, we considered characteristic points of a few observables to explore the QCD 
phase diagram in the $B-T$ plane. This analysis indicates that 
the transition temperature decreases as $B$ grows, in agreement with our previous 
results where other thermodynamic observables (light quark condensate, strange 
quark number susceptibility and Polyakov loop) were used~\cite{Bali:2011qj,Bruckmann:2013oba}. 
Lattice results indicating this tendency have also been obtained using overlap fermions in $N_f=2$ 
QCD~\cite{Bornyakov:2013eya} and in two-color QCD with four equally charged staggered 
quark flavors~\cite{Ilgenfritz:2013ara}. 
The reduction of $T_c(B)$ has been reproduced within the bag model~\cite{Fraga:2012fs} 
and is also supported 
by large $N_c$ arguments~\cite{Fraga:2012ev}. Nevertheless, this feature 
remains a property of the chiral/deconfinement transition 
that most low-energy effective theories or models cannot reproduce, or only for a 
limited range of magnetic fields, see, e.g., Refs.~\cite{Fayazbakhsh:2010bh,Fraga:2013ova,Andersen:2013swa}. 
Recent studies of the Nambu-Jona-Lasinio model, however, indicate that taking into 
account a $B$-dependent Polyakov loop scale parameter~\cite{Ferreira:2013tba}, or the 
magnetic field-induced running of the strong 
coupling~\cite{Farias:2014eca,Ferreira:2014kpa,Ayala:2014iba} might resolve this discrepancy.

\acknowledgments

Our work was supported by the DFG (SFB/TRR 55, BR 2872/6-1), the EU (ITN STRONGnet 238353 
and ERC No 208740) and the Alexander von Humboldt Foundation.
The authors thank Vladimir Braun, Massimo D'Elia, Zolt\'an Fodor, Eduardo Fraga, Igor Shovkovy and K\'alm\'an Szab\'o for useful discussions.

\appendix
\section{Expansion of the quark determinant}
\label{app:expand}

In this appendix we calculate the expansion of the determinant for asymptotically 
large quark masses. 
We consider one quark flavor with electric charge $q$ and mass $m$ and for simplicity, 
set the lattice spacing to unity. 
Using chiral symmetry, the fermionic action can be rewritten as
\be
\log\det M \equiv \log \det [\slashed{D}+m] = \frac{1}{2} \log\det \big[-\slashed{D}^2 + m^2\big] = 
 \frac{1}{2} \,\tr \log \big[1-\slashed{D}^2/m^2\big] + \textmd{const.}
\label{eq:A1}
\ee
Note that the staggered lattice discretization of the Dirac operator only possesses a remnant $\mathrm{U}(1)$ chiral symmetry. This symmetry corresponds to $\{\slashed{D},\eta_5\}=0$, ($\eta_5=(-1)^{n_x+n_y+n_z+n_t}$ is the staggered equivalent of the fifth gamma-matrix) and
allowed us 
to derive Eq.~(\ref{eq:A1}). 
The square of the Dirac operator in the magnetic field is rewritten using $\gamma$-matrix identities as
\be
\slashed{D}^2 = D_\mu D_\mu - \sigma_{xy} qB\, \mathds{1} - \frac{1}{2}\sigma_{\mu\nu}G_{\mu\nu},
\ee
where $\sigma_{\mu\nu}=[\gamma_\mu,\gamma_\nu]/(2i)$ is the spin operator, $G_{\mu\nu} = G_{\mu\nu}^a t^a$ the 
non-Abelian field strength with generators $t^a$, 
and the (Abelian) magnetic field $B$ points in the $z$ direction. 
Now, considering 
the change in the fermionic action due to the magnetic field, and 
expanding the logarithm in $m^{-1}$, we obtain
\be
\begin{split}
\Delta \log\det M &= -\frac{1}{2}\, \Delta \tr \bigg[ \frac{\slashed{D}^2}{m^2} + \frac{\slashed{D}^4}{m^4} + \mathcal{O}(\slashed{D}^6/m^6) \bigg] \\
&= -\frac{1}{2m^4} \, \Delta \tr \bigg[D_\mu D_\mu - \sigma_{xy} qB\, \mathds{1} - \frac{1}{2}\sigma_{\mu\nu}G_{\mu\nu}\bigg]^2 + \mathcal{O}(m^{-6}) \\
&=-\frac{(qB)^2}{2m^4} \cdot 4 N_s^3N_tN_c +\mathcal{O}(m^{-6}).
\end{split}
\ee
Here we used that $\tr \sigma_{xy}=\tr t^a = 0$ such that terms linear in 
$qB$ vanish under the trace, and that $\tr\sigma_{xy}^2 \mathds{1}=4N_s^3N_tN_c$. 
Moreover, terms independent of $B$ cancel in $\Delta \log\det M$. 
Note that at any finite lattice spacing, $\slashed{D}$ is bounded from above by the 
largest possible lattice momentum $\sim a^{-1}$. 

In the large mass limit quarks and gluons decouple from each other, and the gluonic contribution to 
$\log\Z$ becomes independent of the magnetic field. Therefore, we obtain
\be
\Delta \log\Z \xrightarrow{ma\gg1} \Delta \log\det M  \propto -(a^2qB)^2/(ma)^4 + \mathcal{O}((ma)^{-6}),
\label{eq:A4}
\ee
where we reinserted the lattice spacing $a$. 
Eq.~(\ref{eq:A4}) shows that $\Delta p_z$ falls off 
as $(ma)^{-4}$ for large quark masses $m\gg a^{-1}$. Thus we have proven that Eq.~(\ref{eq:pres2}) 
holds in the lattice regularization 
once the upper endpoint of the integral exceeds the lattice scale. 
Accordingly, the derivative with respect to the mass, 
$\Delta \bar\psi\psi$ decays as $(ma)^{-5}$ for large masses, as observed in Sec.~\ref{sec:condbeta}.

\section{Magnetic susceptibility in the HRG model}
\label{app:chiHRG}

In this appendix we calculate $\chi_B$ within the Hadron Resonance Gas model. The EoS 
at nonzero magnetic fields was determined in Ref.~\cite{Endrodi:2013cs} by writing the 
free energy density as an integral over the longitudinal momentum and a sum over Landau-levels, 
both of which can be performed numerically. 
To obtain $\chi_B$, it is instead advantageous to use the proper-time representation~\cite{Schwinger:1951nm}, 
where the expansion of $f$ in the magnetic field can be written down directly. 
For hadrons with electric charge $q$ and spin $s=0,1/2$ or $1$, the energy levels in 
the magnetic field read
\be
E(p_z,k,s_z) = \sqrt{p_z^2 + m^2 + qB(2k+1-2s_z)},
\label{eq:energies}
\ee
where $s_z$ is the projection of the spin on the magnetic field, and we approximated 
the gyromagnetic factor of the hadron as $g=2$. 
To calculate the renormalized susceptibility, it suffices to determine the difference of 
free energies at $T$ and at $T=0$,
\be
f^{s}(T)-f^{s}(0) = (-1)^{2s+1} \frac{qB}{8\pi^2} \int_0^\infty \frac{\dd t}{t^2} e^{-m^2t} \frac{1}{2\sinh(qBt)} \cdot
\left[ \Theta_3 \left( \varphi_s, e^{-1/(4T^2t)} \right) - 1 \right] \cdot \sum_{s_z=-s}^s e^{-2qB s_zt},
\label{eq:fff}
\ee
where the prefactor $(-1)^{2s+1}$ reflects the fermionic/bosonic nature of the hadron, 
the elliptic $\Theta$-function results from summing over Matsubara-frequencies and 
the factor $2\sinh(qBt)$ from summing over the angular momenta $k$ in Eq.~(\ref{eq:energies}).
The first argument of the $\Theta$-function is $\varphi_s=0$ for $s=0,1$ and $\varphi_s=\pi/2$ for 
$s=1/2$, according to the lowest Matsubara-mode. The $-1$ in the square brackets corresponds 
to the subtraction of the $T=0$ term. Note that Eq.~(\ref{eq:fff}) gives the 
contribution of a particle and its antiparticle to the free energy density. 

Keeping quadratic terms in the magnetic field gives the susceptibility,
\be
\chi_B^s(T) = -\left. \frac{\partial^2 [f^s(T)-f^s(0)] }{\partial (eB)^2} \right|_{B=0}= 
(-1)^{2s} \frac{1}{4\pi^2} (q/e)^2 \int_0^\infty \frac{\dd t}{t} e^{-m^2t/T^2}
\left[ \Theta_3\left(\varphi_s, e^{-1/(4t)}\right)-1 \right]  \omega_s,
\ee
where we performed the sum over $s_z$, resulting in the factors
\be
\omega_0 = -1/12, \quad\quad \omega_{1/2} = 1/3, \quad\quad \omega_1 = 7/4,
\ee
and made a change in the integration variable. The remaining integral over $t$ can be performed 
numerically. Inspecting the behavior of the $\Theta$-function we see that $\chi^s_B(T)$ is negative 
for $s=0$ and positive for $s=1/2, 1$: charged pions contribute to diamagnetism (this was 
also recognized in Ref.~\cite{Bonati:2013lca}), whereas protons 
and charged $\rho$-mesons to paramagnetism. 
This tendency can be qualitatively understood invoking the following argument 
(see the discussion in Ref.~\cite{Endrodi:2013cs}). 
The susceptibility is determined by the thermal part of the free energy, which 
contains $\exp(-m(B)/T)$ where $m(B)$ is the effective mass of the hadron at non-zero 
magnetic fields. According to the structure of the lowest Landau-level 
(Eq.~(\ref{eq:energies}) with $p_z=k=0$ and $s_z=s$), for scalar (vector) hadrons this 
effective mass increases (decreases) as $B$ grows. Therefore, the magnitude of the 
thermal free energy is suppressed by the magnetic field for pions, whereas it is 
enhanced for $\rho$-mesons, responsible for the different signs of $\chi_B^s$ in the two cases. 
In the $s=1/2$ case, the lowest level is independent of $B$, thus the sign of 
the susceptibility cannot be anticipated from this argument.
As already mentioned in Sec.~\ref{sec:perme}, at $\mathcal{O}(B^4)$ the vacuum part 
starts to contribute to the free energy as well and turns the magnetization positive 
for each spin channel, given that $eB\gtrsim 0.05 \textmd{ GeV}^2$.

The total susceptibility is obtained as the 
sum over all hadrons
\be
\chi_B(T) = \sum_h d_h \cdot \chi_B^{s_h}(T),
\ee
with multiplicities $d_h$. The list of hadrons taken into account can be found in Ref.~\cite{Endrodi:2013cs}.

\bibliographystyle{jhep}
\bibliography{logZB}

\end{document}